\documentclass[twocolumn,showpacs,preprintnumbers,prb,amsmath,amssymb]{revtex4}
\usepackage{epsfig}
\renewcommand{\theequation}{\arabic{section}.\arabic{equation}}
\begin{document}
%
%
%
%
%
\title{Dynamic scaling in the vicinity of the Luttinger liquid fixed point}

\author{Tom Busche, Lorenz Bartosch and Peter Kopietz}
\affiliation{Institut f\"{u}r Theoretische Physik, Universit\"{a}t Frankfurt,
Robert-Mayer-Strasse 8, 60054 Frankfurt, Germany}
\date{February 7, 2002}
\begin{abstract}
We calculate the single-particle spectral function
$A ( k , \omega )$ of a one-dimensional Luttinger liquid
by means of a functional renormalization group (RG) approach.
Given an infrared energy cutoff $ \Lambda = \Lambda_0 e^{- l}$, our approach
yields the spectral function in the scaling form,
$A_{\Lambda} ( k_F + p , \omega ) =   \tau Z_l
 \tilde{A}_l ( p \xi , \omega \tau )$, where
$k_F$ is the Fermi momentum, 
$Z_l$ is the wave-function renormalization factor,
$\tau = 1 / \Lambda$ is the time scale and $ \xi = v_F / \Lambda$
is the length scale associated with $\Lambda$.
At the Luttinger liquid fixed point ($ l \rightarrow \infty$)
our RG result for $A ( k  , \omega )$
exhibits the correct anomalous scaling properties, and 
for $ k = \pm k_F$ agrees exactly
with the well-known bosonization result at weak coupling.
Our calculation demonstrates 
that the  field rescaling is essential for obtaining
the crossover from Fermi liquid behavior to Luttinger liquid
behavior from a truncation of the hierarchy of exact RG flow equations
as the infrared cutoff $\Lambda$ is reduced.
\end{abstract}
\maketitle
%
%
%
%
\section{Introduction}
\label{sec:intro}

For many years the
renormalization group (RG) 
has been used to study
interacting Fermi systems 
in one spatial dimension ($1d$).\cite{Solyom79}
The success of RG methods in $1d$ relies on the fact that
at low energies, the two-body interactions between fermions
can be parameterized in terms of four coupling constants,
which are usually called $g_1$ (backward scattering),
$g_2$ (forward scattering of fermions propagating in opposite directions),
$g_3$ (Umklapp scattering), and $g_4$ (forward scattering of fermions 
propagating in the same direction).
Most authors have focused on the calculation 
of the RG $\beta$-functions,
which in the Wilsonian RG\cite{Wilson72,Ma76,Fisher98}
describe the 
flow of these couplings as the degrees of freedom are eliminated 
and rescaled.\cite{footnotesolyom}
However, the RG $\beta$-functions do not completely describe the physical 
behavior of the system.
In particular, the RG $\beta$-functions do not contain 
information about the single-particle excitations.
To investigate these, one should calculate
the momentum- and frequency-dependent 
single-particle spectral function
$A (k , \omega )$, which at temperature $T=0$ is related to the
imaginary part of the single-particle Green function 
$G ( k , \omega )$ via 
\begin{equation}
 A  ( k , \omega ) = - \frac{1}{\pi} \,{\rm Im}\, G ( k , \omega + i 0 )
 \; .
 \end{equation}
In the absence of backward and Umklapp scattering, the
leading asymptotic long-distance and long-time behavior of the
single-particle Green function $G ( x , t ) $ in the space-time domain can be
calculated via bosonization if the energy dispersion is
linearized around the Fermi points (Tomonaga-Luttinger model, 
TLM).\cite{Emery79,Voit95}
Obviously, in order to obtain the spectral function,
one should calculate the Fourier-transform  $ G ( k , \omega )$ of $ G (  x, t )$, 
which in general cannot be done exactly.
For the spinless TLM with $g_2$-interactions
a mathematically non-rigorous but physically reasonable
asymptotic analysis yields at temperature $T = 0$
for wave-vectors close to $\pm k_F$ and for low energies,\cite{Luther74,Meden92,Voit93,Meden99}
 \begin{eqnarray}
 A_{\rm TL} (  \pm (k_F + p)  , 
 \omega ) & \approx &     \Lambda_0^{-\eta}   
\frac{\eta}{2} \Theta ( \omega^2 - (v_c p )^2 )
 \nonumber
 \\
 & & \!\!\!\!\!\!\!\!\!\!\!\!\!\!\!\times
 | \omega - v_c p |^{-1 + {\eta}/{2}  } | \omega + v_c  p |^{\eta/2}
 \;  , \quad
 \label{eq:specbos} 
\end{eqnarray} 
where
$ \Lambda_0$ is some ultraviolet cutoff with units of energy (for example
a band width cutoff).
At weak coupling the anomalous dimension $\eta$ and the 
velocity $v_c$ of collective charge excitations are to leading order
 \begin{eqnarray}
 \eta & \approx & \frac{\tilde{g}^2}{8} \;,
 \label{eq:etabos}
 \\ 
 v_c & \approx & v_F \left( 1 - \frac{\tilde{g}^2}{8} \right)
 \label{eq:vcbos}
 \; .
 \end{eqnarray}
Here $v_F$ is the bare Fermi velocity and 
 \begin{equation}
 \tilde{g} = g_2 / ( \pi v_F )
 \label{eq:g2rendef}
 \end{equation}
 is the dimensionless
coupling describing forward scattering of electrons propagating in
different directions.
Eq.\ (\ref{eq:specbos}) 
satisfies a simple scaling law:
for an {\it{arbitrary}} length $\xi$ we may write
  \begin{equation}
 A_{\rm TL} (\pm(k_F + p) , 
 \omega ) = \tau \left( \frac{\xi_0}{\xi } \right)^{  \eta } 
 \tilde{A}_{\rm TL } ( p \xi , \omega \tau ) 
 \;  ,
 \label{eq:specbosscale} 
\end{equation} 
where $\tau =  \xi / v_F$ 
is the time scale associated with $\xi$, and 
the length $
 \xi_0 = v_F / \Lambda_0$ corresponds to the
ultraviolet cutoff $\Lambda_0$.
The dimensionless scaling function $\tilde{A}_{\rm TL } ( q , \epsilon )$  is
 \begin{equation}
 \tilde{A}_{\rm TL} ( q , \epsilon ) = \frac{\eta}{2} \Theta ( \epsilon^2 -
(\tilde{v} q )^2 ) | \epsilon - \tilde{v} q |^{ -1 + \eta/2} | \epsilon + \tilde{v} q |^{\eta / 2}
 \label{eq:tildeAscale}
 \; ,
 \end{equation}
where $ q$ and $\epsilon$ are dimensionless variables, and
$\tilde{v} = v_c / v_F$ is the dimensionless velocity renormalization factor.
Obviously, the dynamic exponent (defined via $\tau \propto \xi^z$) is $z=1$.
The dynamic scaling function (\ref{eq:tildeAscale}) is scale invariant, so that
for an arbitrary scale factor $s$
 \begin{equation}
 \tilde{A}_{\rm TL} ( s q , s \epsilon ) = s^{-1 + \eta }
 \tilde{A}_{\rm TL} ( q , \epsilon )
 \; .
 \label{eq:scaleATL}
 \end{equation}
This scale invariance is a consequence of the fact that 
the TLM 
represents a critical system, corresponding to the
Luttinger liquid fixed point.\cite{Haldane81}
Hence, the 
RG $\beta$-function of the TLM vanishes identically.
It is generally accepted that the TLM
describes the generic low-energy and long-wavelength
properties 
of one-dimensional Fermi systems with dominant forward scattering.
Thus, the TLM is an effective
model which in the regime where backward and Umklapp scattering
are irrelevant emerges at low energies 
when the high-energy degrees of freedom are integrated out
in a Wilsonian RG.

In the theory of dynamic critical
phenomena\cite{Halperin69,Sachdev99}
it is usually assumed that
correlation functions close to the critical point
can be written in the dynamic scaling form
(\ref{eq:specbosscale}), 
with some dynamic scaling function $\tilde{A} ( p \xi , \omega \tau )$.
The dynamic scaling function
depends on $\xi$ only implicitly via 
$ p \xi$ and $ \omega \tau = \omega \xi / v_F$.
For example, at finite temperatures, where we may identify
$\xi = v_F / T$ and $ \tau \approx 1/T$, the scaling function
$\tilde{A}_{\rm TL} ( p \xi , \omega \tau )$ 
of the TLM  has been discussed by
Orgad, Kivelson, and collaborators.\cite{Orgad00}

Suppose now that
we iterate the RG to some 
large scale factor $l = \ln ( \Lambda_0 / \Lambda )$, 
and write the resulting
spectral function  
in the scaling form analogous
to Eq.\ (\ref{eq:specbosscale}),
 \begin{equation}
 A_{\Lambda }  (  k_F + p  , 
 \omega ) = \tau \left( \frac{\xi_0}{\xi } \right)^{  \eta } 
 \tilde{A}_l ( p \xi , \omega \tau ) 
 \;  .
 \label{eq:dynscale} 
\end{equation}
According to the above argument, we would expect that
for $l \rightarrow \infty $ the scaling function
$\tilde{A}_l ( q , \epsilon )$  becomes
independent of $l$  and
approaches $ \tilde{A}_{\rm TL} ( q , \epsilon )$ for large
$| q |$ and $ | \epsilon |$.
In this work we shall show that this 
expectation is not quite correct: even if both 
$ | q |$ and $|\epsilon|$ are
large compared with unity, the difference $|\epsilon| - \tilde v |q|$ can be small.
In this regime we find that
 $\tilde{A}_\infty ( q , \epsilon ) = \lim_{l \rightarrow \infty} 
\tilde{A}_l ( q , \epsilon )$ behaves quite differently from the 
scaling function of the TLM.
Physically, this is due to the fact that
the spectral line-shape 
of a generic one-dimensional Fermi system with
dominant forward scattering exhibits in general some non-universal features
which are determined by irrelevant couplings and
thus spoil the universality of the spectral function.
In fact, even the spectral function of the TLM exhibits some non-universal features.\cite{Meden99}

We shall here
attempt to calculate the 
spectral function of the $g_2$-TLM  
using RG methods. Therefore 
we shall calculate  the RG flow of the 
irreducible {\it{two-point}} vertex.
Recall that
the usual RG $\beta$-function describes the flow of the 
momentum- and frequency-independent part of the {\it{four-point}} vertex.
Surprisingly,  the problem of calculating the spectral function of 
a strongly correlated fermionic system like the 
TLM via RG methods 
has not received much attention.
In fact, with the exception of the
recent work by Ferraz,\cite{Ferraz01} where
the spectral function of a special
$2d$ Fermi system  has been calculated
by means of the field theory RG, 
we are not aware of any RG calculation of the spectral line shape
of a strongly correlated fermionic many-body system.
Note that in order
to obtain a non-trivial $k$- and $\omega$-dependence and the anomalous scaling properties
given in Eq.\ (\ref{eq:specbosscale}), one should retain
infinitely many couplings which are irrelevant by naive power counting.
In other words, one needs to keep track of the RG flow of  coupling functions.
At the first sight, this seems to be a formidable task, which is impossible
to carry out in practice.
 Nevertheless, in this work we shall show 
that  at weak coupling a simple truncation of the hierarchy of
functional RG equations
for the irreducible vertex functions  
yields an expression for the spectral function which 
has the correct anomalous scaling
properties and, at  least for $k = \pm k_F$, agrees with
the exact result known from bosonization.

\section{Exact flow equations}
\label{sec:rescale}
\setcounter{equation}{0}

Originally, exact RG flow equations have been
developed in field theory and statistical physics
to study systems in the vicinity of a critical 
point.\cite{Wegner73,Polchinski84,Nicoll76,Wetterich93,Morris94}
Recently, several authors have started to apply
these methods to interacting Fermi systems at finite 
densities.\cite{Zanchi96,Salmhofer98,Halboth00}
In the normal state the existence of a Fermi surface
adds some new complexity to the problem: 
because the single particle Green function
exhibits singularities on the entire Fermi surface,
the Fermi surface plays the role of a continuum (in $d > 1$) 
of RG fixed points.
However, this fixed point manifold
is not known a priori, and should be calculated self-consistently
within the RG. In
Ref.\ \onlinecite{Kopietz01b} we have shown how 
the true Fermi surface of the interacting system can
be calculated from the  
requirement that the RG
approaches a fixed point. 
Fortunately,
in $1d$ we know a priori that the Fermi surface (which consists of
two points $\pm k_F$) is not renormalized by the 
interactions,\cite{Blagoev97}  
so that the above mentioned problem of a self-consistent
determination of the Fermi surface does not arise.

Even if the shape of the Fermi surface is known,
the implementation of the exact RG for
Fermi systems is non-trivial and some open problems remain.
In particular, in Ref.\ \onlinecite{Kopietz01b} we have pointed out
that the RG approach  proposed in Refs.\ \onlinecite{Zanchi96,Salmhofer98,Halboth00}
includes only the mode-elimination step, but
completely omits the usual rescaling.
We have further speculated that this incomplete RG transformation
 might be responsible for the runaway flow to strong coupling which
is typically encountered in these works.
The purpose of this article is to further substantiate this claim
by applying the functional RG method to a non-trivial yet exactly solvable
many-body problem where the results can be critically compared
with known results obtained by other methods.

\subsection{Definition of the model and notations}
\label{subsec:scaling}

Starting point of our calculation are the exact flow equations 
derived in Ref.\ \onlinecite{Kopietz01b}, which give the RG flow
of the one-particle irreducible 
vertices when degrees of freedom are integrated out
and a subsequent rescaling step is applied.
For convenience we shall now briefly 
present these equations for the special case of $1d$,
using a slightly different notation than in Ref.\ \onlinecite{Kopietz01b}.
We shall work at zero temperature, but expect that
our results also describe  temperatures $T > 0$ as long as we consider energy
scales large compared with $T$.
In $1d$, the Fermi surface consists of two points,
$\alpha k_F$,  where  $\alpha = \pm 1$  labels the right and left Fermi point.
An arbitrary wave-vector $k$ can be written as
  \begin{equation}
 k = \alpha ( k_F +  p )
 \; , 
\label{eq:trans}
 \end{equation}
where $\alpha = +1$ for $k > 0$ and $\alpha = -1$ for $k <0$.
Note that the deviation $p = \alpha (k - \alpha k_F)$
from $\pm k_F$ is always measured locally outwards, which
is different from the convention usually  
adopted in bosonization.\cite{Solyom79,Emery79,Voit95,Haldane81}
For small $p$ we may expand the non-interacting
energy dispersion $ \epsilon_{k}$
around the Fermi points,
 \begin{equation}
 \epsilon_{ \alpha (k_F + p ) } =  \epsilon_{  k_F } +
v_F p + 
\frac{ p^2 }{2 m }
 + \ldots
 \; .
 \label{eq:epsexpansion}
 \end{equation}
Here $m$ is the bare mass of the fermions.
As discussed in Ref.\ \onlinecite{Kopietz01b},
the expansion (\ref{eq:epsexpansion})
is around the true Fermi momentum of the interacting
many-body system. Of course, in $1d$ the interacting and non-interacting
$k_F$ are identical if we compare systems with the same density.
Note that by time reversal invariance $ \epsilon_{k } = \epsilon_{-k}$, so that
$v_F$ and $m$ are independent of $\alpha$.
For $ | p | \ll k_F$ it is reasonable to linearize the energy dispersion and neglect the term of order $p^2$
in Eq.\ (\ref{eq:epsexpansion}).
Introducing an infrared cutoff $\Lambda$ and an ultraviolet cutoff $\Lambda_0$
with units of energy,
the non-interacting Matsubara Green function is 
 \begin{equation}
 G^{0}_{ \Lambda , \Lambda_0 } ( \alpha ( k_F + p ) , i \omega )
 = \frac{ \Theta ( \Lambda_0 >  v_F | p | > \Lambda ) }{ i \omega - v_F p }
 \label{eq:G0cutoff} 
\; ,
 \end{equation}
where $i \omega$ is a fermionic  Matsubara frequency, and 
\begin{eqnarray}
 \Theta ( x_2 > x > x_1 ) & = & \Theta ( x_2 - x ) - \Theta ( x_1 - x ) 
 \nonumber
 \\
 & = &
 \left\{
 \begin{array}{lc}
 1 & \mbox{if $ x_2 > x > x_1 $} \\
 0 & \mbox{else}
 \end{array}
 \right.
 \; .
 \end{eqnarray}
Later we shall also write
 \begin{equation}
 \Theta ( x_2  > x_1 ) = \Theta ( x_2 - x_1 ) =
 \left\{
 \begin{array}{lc}
 1 & \mbox{if $ x_2 >  x_1 $} \\
 0 & \mbox{else}
 \end{array}
 \right.
 \; .
 \end{equation}
Taking the limit $\Lambda_0 \rightarrow \infty$
in Eq.\ (\ref{eq:G0cutoff}) amounts to extending
the linear energy dispersion on both branches from $- \infty$ to $+ \infty$.
The unphysical states with energies 
far below the Fermi energy
introduced in this way are occupied according to the Pauli principle. 
It is generally accepted that this  filled Dirac sea
is dynamically irrelevant and
does  not modify the low-energy physics.
However, the precise way in which the cutoff is removed
can affect the numerical value of the various 
Luttinger liquid parameters.\cite{Schulz98}
At this point  we shall keep 
$\Lambda_0$ finite and assume that the bare electron-electron interaction
of the model is characterized by a 
momentum- and frequency-independent
totally antisymmetric
irreducible four-point vertex of the form
 \begin{equation}
 \Gamma^{(4)}_{\Lambda_0} ( K_1^{\prime} , K_2^{\prime} ;
 K_2 , K_1 ) =  A_{ \alpha_1^{\prime}  \alpha_2^{\prime} ;
 \alpha_2  \alpha_1 } 
g_0
  \; ,
 \label{eq:g2def}
 \end{equation}
where $K = ( k , i \omega ) = ( \alpha ( k_F + p ) , i \omega )$, and
 \begin{eqnarray}
 A_{ \alpha_1^{\prime}  \alpha_2^{\prime} ;
 \alpha_2  \alpha_1 } & = & \delta_{ \alpha_1^{\prime} , \alpha_1} 
 \delta_{\alpha_2^{\prime} , \alpha_2 } -
 \delta_{ \alpha_2^{\prime} , \alpha_1} 
 \delta_{\alpha_1^{\prime} , \alpha_2 }
 \nonumber
 \\
 & = &
 D_{ \alpha_1^{\prime}  \alpha_2^{\prime} ;
 \alpha_2  \alpha_1 } - E_{ \alpha_1^{\prime}  \alpha_2^{\prime} ;
 \alpha_2  \alpha_1 }
 \label{eq:Adef}
 \end{eqnarray}
is antisymmetric with respect to the exchange
$\alpha_1 \leftrightarrow \alpha_2$ or 
$\alpha_1^{\prime} \leftrightarrow \alpha_2^{\prime}$.
For later convenience we have introduced the notations
 \begin{eqnarray}
   D_{ \alpha_1^{\prime}  \alpha_2^{\prime} ;
 \alpha_2  \alpha_1 } & = & 
 \delta_{ \alpha_1^{\prime} , \alpha_1} 
 \delta_{\alpha_2^{\prime} , \alpha_2 } \;,\; \mbox{(direct term)}
 \\
 E_{ \alpha_1^{\prime}  \alpha_2^{\prime} ;
 \alpha_2  \alpha_1 } & = & 
 \delta_{ \alpha_2^{\prime} , \alpha_1} 
 \delta_{\alpha_1^{\prime} , \alpha_2 } \;,\; \mbox{(exchange term)}
 \; . \quad
 \label{eq:DEdef}
 \end{eqnarray}
In the limits $\Lambda_{0} \rightarrow \infty$ and $\Lambda \rightarrow 0$ 
the Green function of the model defined by Eqs.\ (\ref{eq:G0cutoff}) and 
(\ref{eq:g2def}) can be calculated exactly  in the
space-time domain  via bosonization  (TLM with $g_2$-interactions).
The generally accepted result for the spectral function 
is given in Eq.\ (\ref{eq:specbos}).\cite{footnote:g}

\subsection{Scaling functions and flow equations}

Suppose now that we reduce the infrared cutoff by setting
 \begin{equation}
 \Lambda = \Lambda_0 e^{- l}
 \end{equation}
and follow the evolution of the correlation functions as the
logarithmic flow parameter $l $ increases.
Noting that the infrared cutoff $\Lambda$ has units of energy, 
we may define a length scale
$\xi$ and an associated time scale $\tau$,
 \begin{equation}
 \xi = v_F / \Lambda
 \; \; , \; \;  \tau = 1/ \Lambda = \xi / v_F
 \; .
 \label{eq:xitaudef}
 \end{equation}
From these quantities, we may construct dimensionless
scaling variables,\cite{footnotevariables}
 \begin{eqnarray}
 q = p \xi & = & ( \alpha k -  k_F ) \xi 
 \; \; , \; \; 
\epsilon  =  \omega \tau = \omega/ \Lambda
 \; , \quad
\end{eqnarray}
and write the exact single-particle Matsubara Green function of the theory with 
infrared cutoff $\Lambda$ and ultraviolet cutoff 
$\Lambda_0$ in the dynamic scaling form\cite{Halperin69,Sachdev99}
  \begin{equation}
 G_{\Lambda , \Lambda_0} (  k  , i \omega  )
 = \tau  Z_l   
 \tilde{G}_l  \left(    p  \xi   ,  i \omega \tau  \right)
 \; ,
 \label{eq:Gtscale}
 \end{equation}
where $ \tilde{G}_l ( q , i \epsilon )$ is a dimensionless dynamic scaling function.
Here
the wave-function renormalization factor $Z_l$  is
related to the irreducible self-energy $\Sigma_{\Lambda} ( k , i \omega )$
as usual,
 \begin{equation}
 Z_l = \frac{1}{ 1 - 
 \left. \frac{ \partial \Sigma_{\Lambda} ( k_F , i \omega ) }{ \partial ( i \omega )}
 \right|_{ \omega = 0 } }
 \; .
 \label{eq:Zdef}
 \end{equation}
Note that by time-reversal symmetry $Z_l$ and $\Sigma_{\Lambda}$ are independent of $\alpha$.
It is also convenient to set
 \begin{eqnarray}
 \tilde{G}_l ( q , i \epsilon ) 
 & = &
 \frac{  
 \Theta ( e^l >  | q |  > 1 )}{ 
 r_l ( q , i \epsilon ) }
 \; ,
 \label{eq:Gtres}
\end{eqnarray}
where
 \begin{eqnarray}
  r_l ( q , i \epsilon ) 
 & = & 
  Z_l  \left[ i \epsilon - 
 q  \right] + 
 \tilde{\Gamma}_l^{(2)} ( q , i \epsilon )
 \label{eq:rtdef}
 \; ,
 \end{eqnarray}
and
 \begin{equation}
  \tilde{\Gamma}_{l}^{(2)}
 ( q , i \epsilon  )  = - \tau Z_l 
 \left[ \Sigma_{\Lambda } ( k_F + q/\xi  , i \epsilon / \tau ) - 
 \Sigma ( k_F , i0 ) \right]
 \; .
 \label{eq:Gammatdef}
 \end{equation}
Here $\Sigma ( k , i \omega ) = \lim_{\Lambda \rightarrow 0 } 
\Sigma_{\Lambda} ( k , i \omega )$ 
is the exact self-energy of the model without infrared cutoff.
Following Ref.\ \onlinecite{Kopietz01b}, we also define the scaling functions
for the higher order irreducible vertices,
\begin{eqnarray}
\tilde{\Gamma}_{l}^{ (2n) }
(  Q_1^{\prime} , \ldots , 
Q_n^{\prime} ;  Q_n , \ldots ,  Q_1 ) & = & 
 \nonumber
 \\
& & \hspace{-50mm}
\nu_0^{n-1} \Lambda^{n-2}
Z_l^n
\Gamma^{(2n)}_{\Lambda} ( K_1^{\prime} , \ldots , K_n^{\prime} 
; K_n , \ldots , K_1 )
 \; , \qquad
 \label{eq:Gammarescale}
\end{eqnarray}
where 
$
\Gamma^{(2n)}_{\Lambda} ( K_1^{\prime} , \ldots , K_n^{\prime} 
; K_n , \ldots , K_1 )$ are the usual one-particle irreducible
$2n$-point vertices,
$\nu_0 = (  \pi v_F )^{-1}$ is the density of states of non-interacting
spinless electrons in $1d$, and  $K = ( k , i \omega )$  and
$Q = ( \alpha, q , i \epsilon) $ are composite labels.

The rescaled irreducible two-point vertex defined in Eq.\ (\ref{eq:Gammatdef}) satisfies 
the following exact flow equation,\cite{Kopietz01b}
 \begin{eqnarray}
 \partial_l \tilde{\Gamma}_l^{(2)} ( Q ) & = &
 ( 1 - \eta_l - Q \cdot \partial_Q  )
 \tilde{\Gamma}_l^{(2)} ( Q )
 + \dot{\Gamma}_l^{(2)} ( Q )
 \; , \qquad \quad
 \label{eq:twopoint}
 \end{eqnarray}
where
 \begin{eqnarray}
   \dot{\Gamma}_l^{(2)} ( Q ) 
 & =  & - \frac{1}{2} \sum_{\alpha^{\prime}}  \int d q^{\prime} \int 
 \frac{ d \epsilon^{\prime}}{2 \pi}\,
 \dot{G}_l ( Q^{\prime} ) 
 \nonumber
 \\
 & \times &
 \tilde{\Gamma}_{l}^{(4)}
 (  Q   , Q^{\prime}
 ;  Q^{\prime} ,  Q )
 \; ,
 \label{eq:B2def}
 \end{eqnarray}
and $\eta_l$ is the flowing anomalous dimension,
 \begin{equation}
 \eta_l = - \partial_l \ln Z_l =- \frac{ \partial_l Z_l}{Z_l}
 \; .
 \label{eq:eta_l}
 \end{equation}
We have introduced the notations
 \begin{equation}
 Q \cdot \partial_Q
 = q \partial_q + \epsilon \partial_{\epsilon}
 \; ,
 \end{equation}
 \begin{equation}
 \dot{{G}}_l  ( Q ) = \frac{ \delta ( | q |  -1 )}{
 r_l ( Q )}
 \; .
 \end{equation}
Eq.\ (\ref{eq:twopoint}) is equivalent with the following integral equation,
  \begin{eqnarray}
 \tilde{\Gamma}_l^{(2)} ( Q ) & = & 
 \tilde{\Gamma}_{l=0}^{(2)} ( Q ) 
 \nonumber
 \\
 & + & \int_0^l d t \, e^{ t  - \bar{\eta}_l ( t ) }
 ( 1 - \eta_{l- t }  )  \tilde{\Gamma}_{l=0}^{(2)} ( e^{- t} Q ) 
 \nonumber
 \\
 & + & \int_0^l d t \, e^{ t  - \bar{\eta}_{l} ( t )  }
 \dot{\Gamma}^{(2)}_{ l - t } ( e^{-t } Q )
 \; ,
 \label{eq:twopointint}
 \end{eqnarray}
where $e^{-t}  Q = ( \alpha , e^{-t} q , e^{-t} i \epsilon )$, and 
we have defined
 \begin{equation}
 \bar{\eta}_{l} ( t ) =  \int_{ 0}^{t} 
d t^{\prime}  \eta_{l - t + t^{\prime } } 
 \; .
 \end{equation}
The flow of the inhomogeneity $\dot{\Gamma}_l^{(2)} ( Q )$
on the right-hand side of Eq.\ (\ref{eq:twopoint}) is controlled by
the scaling function of the four-point vertex $\tilde{\Gamma}_{l}^{ (4)}$, 
which satisfies the exact flow equation 
\begin{widetext}
 \begin{eqnarray}
 \partial_l \tilde{\Gamma}_{l}^{ (4)}
 (   Q_1^{\prime} ,   Q_2^{\prime}; 
 Q_2 ,  Q_1 ) = 
 - \left[ 2 \eta_l + \sum_{i = 1}^{2}  \left(
 Q_i^{\prime} \cdot \partial_{Q_i^{\prime}}  + 
 Q_i \cdot \partial_{Q_i} \right)
 \right] 
 \tilde{\Gamma}_{l}^{(4) }
 (   Q_1^{\prime} ,   Q_2^{\prime}; 
   Q_2 ,  Q_1 ) +
  \dot{\Gamma}_{l}^{(4) }
  (   Q_1^{\prime} ,   Q_2^{\prime}; 
   Q_2 ,  Q_1 ) 
 \label{eq:fourpoint}
 \; ,
 \end{eqnarray}
where
 \begin{eqnarray}
 \dot{\Gamma}_{l}^{(4) }
  (   Q_1^{\prime} ,  Q_2^{\prime}; 
    Q_2 ,  Q_1 ) & = &
    -   \frac{1}{2} \sum_{\alpha} \int dq \int \frac{ d \epsilon}{ 2 \pi }\,
 \dot{G}_l (   Q  )
  \tilde{\Gamma}_{l}^{(6)}
 (   Q_1^{\prime},    
 Q_2^{\prime} ,
   Q ;   Q ,  Q_2 ,  Q_1 )
 \nonumber
 \\
 &  & \hspace{-2mm} {} -  
 \frac{1}{2} \sum_{\alpha} \int dq \int \frac{ d \epsilon}{ 2 \pi } 
 \left[ \dot{G}_l (  Q  )  \tilde{G}_l 
(  Q^{\prime} )  + 
 \tilde{G}_l ( Q )  \dot{G}_l (  Q^{\prime} )
 \right]
 \nonumber
 \\
 &  & \hspace{-2mm} {} \times \left\{ \frac{1}{2}
 \left[
 \tilde{\Gamma}_{l}^{(4)}
 (  Q_1^{\prime} ,   Q_2^{\prime}; 
  Q^{\prime} ,   Q )
 \tilde{\Gamma}_{l}^{ (4) }
 (   Q ,  Q^{\prime} ;   Q_2 ,  Q_1)
 \right]_{ K^{\prime} = K_1 + K_2 - K}
 \right.
 \nonumber
 \\
 &  & 
 \hspace{-2mm} {} -  \left[
 \tilde{\Gamma}_{l}^{ (4) }
 (   Q_1^{\prime},   Q^{\prime} ; 
 Q  ,   Q_1 )
 \tilde{\Gamma}_{l}^{(4) }
 (   Q_2^{\prime},   Q  ;   Q^{\prime}
  ,  Q_2 )
 \right]_{ K^{\prime} = K + K_1 - K_1^{\prime}}
 \nonumber
 \\
 &  &
 \left.
 \hspace{-2mm} {}
 + 
\left[
 \tilde{\Gamma}_{l}^{(4) }
 (  Q_2^{\prime} ,  Q^{\prime} ;  Q  ,   Q_1 ) 
 \tilde{\Gamma}_{l}^{(4) }
 (  Q_1^{\prime},   Q ;  Q^{\prime},  Q_2 )
 \right]_{ K^{\prime} = K + K_1 - K_2^{\prime}}
 \right\} \; .
 \label{eq:B4def}
 \end{eqnarray}
Here $\tilde{\Gamma}^{(6)}_l$ is the dimensionless irreducible six-point vertex.
As Eq.\ (\ref{eq:B2def}) can be converted into Eq.\ (\ref{eq:twopointint}),
Eq.\ (\ref{eq:fourpoint}) can be transformed into the integral equation
 \begin{eqnarray}
  \tilde{\Gamma}_{l}^{ (4)}
 (  Q_1^{\prime} ,  Q_2^{\prime}; 
  Q_2 ,  Q_1 ) 
 & = &
  \tilde{\Gamma}_{l=0}^{ (4)}
  (   Q_1^{\prime} ,  Q_2^{\prime}; 
   Q_2 ,  Q_1 ) 
   -  \int_{0}^{l} d t e^{ - 2 \bar{\eta}_{l} ( t )   } 2 \eta_{ l- t }
  \tilde{\Gamma}_{l=0}^{ (4)}
  (  e^{- t } Q_1^{\prime} ,   
 e^{- t} Q_2^{\prime}; 
   e^{- t } Q_2 ,  e^{- t } Q_1 )
 \nonumber
 \\
 &  & \hspace{10mm} +  
 \int_{0}^{l} d t e^{ - 2 \bar{\eta}_{l} ( t ) }
  \dot{\Gamma}^{(4)}_{l - t }
  (   e^{- t } Q_1^{\prime} ,   
 e^{- t} Q_2^{\prime}; 
   e^{- t } Q_2 , e^{- t } Q_1 ) \; .
 \label{eq:Gamma4integral}
 \end{eqnarray}
\end{widetext}

\subsection{Classification of couplings in the hydrodynamic regime}

The above functional RG equations determine the RG flow of 
the momentum- and frequency-dependent irreducible vertices
$\tilde{\Gamma}^{(2n)}_l  ( Q_1^{\prime} ,
\ldots , Q_n^{\prime}; Q_n , \ldots , Q_1 ) $.
By expanding these vertices  in powers of the 
dimensionless scaling variables $q_i = p_i \xi$ and $\epsilon_i = \omega_i \tau$,
we obtain the RG flow of the coupling constants 
of the model, which are the coefficients in this
multi-dimensional Taylor expansion. 
Following the usual terminology, couplings with positive scaling 
dimension are called relevant,  couplings with  vanishing scaling dimension are called
marginal, and  couplings with negative scaling dimension are called irrelevant.
The irrelevant couplings grow at short distances (i.e. in the {\it{ultraviolet}})
and spoil the renormalizabiliy  of the theory. 
In this case the usual field theory RG
is not applicable. 
On the other hand, in a Wilsonian RG the irrelevant couplings can be treated
on the same footing as the relevant ones.
For this reason the Wilsonian RG is very popular  
in condensed matter physics, where one usually has a physical
ultraviolet cutoff and renormalizability is not necessary. 

What is the regime of validity of the above
expansion? Obviously, the expansion of the 
rescaled vertex functions to some finite order
in powers of the dimensionless scaling variables 
 $q_i = p_i \xi$ and $\epsilon_i = \omega_i \tau$ 
can be justified if these variables are small compared with unity.
As long as the infrared cutoff
$ \Lambda = v_F / \xi = 1 / \tau$ is finite, this condition can be satisfied for sufficiently small
wave-vectors and frequencies. If we identify $\xi$ with the correlation length and
$\tau$ with the corresponding characteristic time scale, then 
the expansion of the
 $\tilde{\Gamma}^{(2n)}_l  ( Q_1^{\prime} ,
\ldots , Q_n^{\prime}; Q_n , \ldots , Q_1 ) $ in powers of the 
scaling variables $q_i = p_i \xi$ and $\epsilon_i = \omega_i \tau$
can be viewed as an expansion of the dynamic scaling functions
in the {\it{hydrodynamic regime}}, where one is interested in length scales larger 
than $\xi$ and time scales longer than $\tau$. 
Of course, a normal metal at zero temperature is a critical system, where
$\xi$ and $\tau$ diverge. 
Close to this critical state (where
$\xi$ and $\tau$ are large but finite), we expect that the correlation functions
assume some scaling form.
This  corresponds to the {\it{critical regime}}
(also called {\it{scaling regime}}).
For example,
we may describe a normal metal at finite temperature
in terms of flow equations derived for $T=0$ if we 
stop scaling the flow equations at a finite length scale
of the order of $\xi \approx v_F / T$. 

For certain systems it may happen that
the couplings defined in the hydrodynamic regime
 smoothly evolve
into analogous couplings in the scaling regime, where both
$q_i$ and $\epsilon_i$ are large.
This is the essence of the dynamic scaling hypothesis.\cite{Halperin69}
For example, 
for a Fermi liquid we expect that 
such a smooth crossover between the hydrodynamic and the scaling regime indeed exists, 
because 
by definition the  self-energy is analytic 
and hence can be expanded in powers of momenta and frequencies.
On the other hand,  the two-point function of a Luttinger liquid
is known to exhibit algebraic singularities, so that
there cannot be a smooth connection between the hydrodynamic and 
the critical regime. In this case one expects the analytic properties of the dynamic scaling
function $\tilde{\Gamma}^{(2)}_l ( q , i \epsilon )$ to change as we 
move from the hydrodynamic into the
scaling regime. In this work we shall show that this is indeed the case and
give an approximate expression for the scaling function. 

Let us now classify the couplings according to their relevance
in the hydrodynamic regime. 
First of all, the momentum- and frequency-independent part of
the two-point vertex,
 \begin{equation}
 \tilde{\mu}_l  =  \tilde{\Gamma}^{(2)}_l ( 0 )
 \; ,
 \label{eq:mutdef}
 \end{equation}
is relevant. 
Here $  \tilde{\Gamma}^{(2)}_l ( 0 ) $ 
stands for $\tilde{\Gamma}^{(2)}_l ( \alpha ,  q = 0 , i \epsilon = i0 )$.
From Eq.\ (\ref{eq:twopoint}) it is easy to see that
$\tilde{\mu}_l$ satisfies the exact flow equation
 \begin{equation}
 \partial_l \tilde{\mu}_l = ( 1 - \eta_l ) \tilde{\mu}_l + \dot{\Gamma}^{(2)}_l ( 0 )
 \; .
 \label{eq:muflow}
 \end{equation}
Note that in general we have to fine-tune the other couplings such that
the relevant coupling $\tilde{\mu}_l$ approaches a fixed point. 
The fixed point value of $\tilde{\mu}_l$ for $ l \rightarrow \infty$ is then determined by
 \begin{equation}
 ( 1 -  \eta_{\infty} ) \tilde{\mu}_\infty =    -  \dot{\Gamma}^{(2)}_{\infty} ( 0 )
 \; .
 \label{eq:mufixed}
 \end{equation}
Because $ \dot{\Gamma}^{(2)}_{\infty} ( 0 )$ implicitly
depends on $\tilde{\mu}_{\infty}$,  
Eq.\  (\ref{eq:mufixed}) is really a self-consistency equation for
the fixed point value of $\tilde{\mu}_l$.

There are
two marginal couplings associated with
the irreducible two-point vertex 
$\tilde{\Gamma}^{(2)}_l ( Q )$:
the wave-function renormalization factor $Z_l$, and 
the Fermi velocity renormalization factor $\tilde{v}_l$.
Using the definition  (\ref{eq:Zdef}), 
it is easy to show that 
$Z_l$ can be expressed  in terms of our rescaled  two-point vertex as follows,
 \begin{equation}
 Z_l = 1 -  \left.
\frac{ \partial \tilde{\Gamma}^{(2)}_l ( \alpha , 0 , i \epsilon )}{ \partial ( i \epsilon )}
 \right|_{ \epsilon = 0 }
 \label{eq:Zgamma2}
 \; .
 \end{equation}
The dimensionless Fermi velocity renormalization factor
can be written as
  \begin{equation}
 \tilde{v}_l = Z_l -  \left.
\frac{ \partial \tilde{\Gamma}^{(2)}_l ( \alpha , q , i 0 )}{ \partial q}
 \right|_{ q = 0 }
 \label{eq:vtdef}
 \; .
 \end{equation}
Note that for a Fermi liquid
$\tilde{v}_l$  can be identified with the dimensionless inverse effective mass renormalization,
 \begin{equation}
 \tilde{v}_l = \frac{ m}{m^{\ast}_l }
 \; ,
 \label{eq:meffren}
 \end{equation}
where $m$ is the bare mass and $m^{\ast}_l$ is the effective mass of the model
with infrared cutoff $ \Lambda_0 e^{-l}$. 
The  expansion of the scaling function associated with the two-point vertex for small
$q$ and $\epsilon$ then reads
 \begin{eqnarray}
 \tilde{\Gamma}^{(2)}_l ( Q ) & = & \tilde{\mu}_l 
 + ( 1 - Z_l ) i \epsilon + ( Z_l - \tilde{v}_l ) q 
 \nonumber
 \\
 & & {}
+ {\mathcal O} ( q^2,  \epsilon^2 , q \epsilon  )
 \; ,
 \label{eq:Gamma2expansion}
 \end{eqnarray}
so that for small $q$ and $\epsilon$
the dimensionless inverse propagator defined in Eq.\ (\ref{eq:rtdef})
is given by 
 \begin{eqnarray}
 r_l ( Q ) & = &
  Z_l  \left[ i \epsilon -  q \right] + 
 \tilde{\Gamma}_l^{(2)} ( Q )
 \nonumber
 \\
 & = & i \epsilon - \tilde{v}_l q +  \tilde{\mu}_l + {\mathcal O} ( q^2, \epsilon^2 ,  q \epsilon  )
 \; .
 \label{eq:rtexpansion}
 \end{eqnarray}
The exact flow equations for the marginal couplings
$Z_l$ and $\tilde{v}_l$ are easily obtained from Eqs.\ (\ref{eq:twopoint}), (\ref{eq:eta_l}) and (\ref{eq:vtdef}),
 \begin{equation}
 \partial_l Z_l = - \eta_l Z_l
 \label{eq:Ztflow}
 \; ,
 \end{equation}
  \begin{equation}
 \partial_l \tilde{v}_l = - \eta_l \tilde{v}_l - 
 \left. \frac{ \partial \dot{\Gamma}_l^{(2)} ( \alpha , q , i 0 ) }{ \partial q } \right|_{q = 0 }
 \label{eq:vtflow}
 \; .
 \end{equation}
Differentiating Eq.\ (\ref{eq:twopoint}) with respect to $i\epsilon$, setting $\epsilon = 0$, and using Eq.\ (\ref{eq:Zgamma2}), we see that the flowing 
anomalous dimension $\eta_l$ is related to the function $\dot{\Gamma}^{(2)}_l ( Q ) $ via
 \begin{equation}
 \eta_l = 
  \left. \frac{ \partial \dot{\Gamma}_l^{(2)} ( \alpha , 0 , 
 i \epsilon ) }{ \partial ( i \epsilon) } \right|_{\epsilon = 0 }
 \; .
 \label{eq:etatB}
 \end{equation}
Since
$\dot{\Gamma}_l^{(2)} ( \alpha , 0 , 
 i \epsilon )$ is defined in terms of the four-point vertex $\tilde{\Gamma}_{l}^{(4)}$ 
via Eq.\ (\ref{eq:B2def}), we can obtain
the following explicit expression for the flowing anomalous dimension,
 \begin{eqnarray}
  \eta_l 
 & =  & - \frac{1}{2} \sum_{\alpha^{\prime}}  \int d q^{\prime} \int 
 \frac{ d \epsilon^{\prime}}{2 \pi}\,
 \dot{G}_l ( Q^{\prime} )
 \nonumber
 \\
 &  \times &  
  \left. \frac{ \partial 
\tilde{\Gamma}_{l}^{(4)}
 ( \alpha , 0, i \epsilon  , Q^{\prime}
 ; Q^{\prime} , \alpha , 0 , i \epsilon ) }{ \partial ( i \epsilon ) } 
 \right|_{ \epsilon = 0 }
 \; .
 \label{eq:anomalexplicit}
 \end{eqnarray}
This expression is also valid in dimensions $d > 1$ provided
the discrete index $\alpha$ is replaced by a $d$-dimensional  unit vector  which
labels the points on the Fermi surface. Thus, 
for a given irreducible four-point vertex, 
Eq.\ (\ref{eq:anomalexplicit}) can be used
to directly calculate the anomalous dimension
of any normal Fermi system.

There is one more marginal coupling, namely the
irreducible four-point vertex at vanishing momenta and frequencies.
Because by construction 
$ \tilde{\Gamma}_{l}^{ (4)}
 ( Q_1^{\prime} , Q_2^{\prime};  Q_2 , Q_1 ) $
 is antisymmetric
with respect to the permutation of the incoming or the outgoing 
fermions, for vanishing external momenta and frequencies
the four-point vertex must be of the form
 \begin{equation}
 \tilde{\Gamma}_{l}^{ (4)}
  ( \alpha_1^{\prime} , 0 , \alpha_2^{\prime} , 0; 
 \alpha_2 , 0 , \alpha_1, 0 )   = A_{ \alpha_1^{\prime}  \alpha_2^{\prime} ;
 \alpha_2  \alpha_1 } 
\tilde{g}_l
  \; ,
 \label{eq:gtdef}
 \end{equation}
where
$
 A_{ \alpha_1^{\prime}  \alpha_2^{\prime} ;
 \alpha_2  \alpha_1 }$ is defined in Eq.\ (\ref{eq:Adef}). 
Hence, the marginal part of the four-point vertex can be
expressed in terms of a single marginal coupling $\tilde{g}_l$, which satisfies
the exact flow equation
 \begin{equation}
 \partial_l \tilde{g}_l = - 2 \eta_l \tilde{g}_l  + B_l
 \; .
 \label{eq:gtflow}
 \end{equation}
Here $B_l$ is defined by 
 \begin{equation}
 \dot{\Gamma}^{(4)}_{l} 
  ( \alpha_1^{\prime} , 0 , \alpha_2^{\prime} , 0 ; 
  \alpha_2 , 0 , \alpha_1, 0 )  =  
 A_{ \alpha_1^{\prime}  \alpha_2^{\prime} ;
 \alpha_2  \alpha_1 } 
 {B}_l
 \; ,
 \label{eq:Btdef}
 \end{equation}
with  $\dot{\Gamma}^{(4)}_{l}$ given  in Eq.\ (\ref{eq:B4def}).

It is convenient to subtract 
the relevant and marginal couplings 
from the exact vertices,
thus defining coupling functions containing only irrelevant couplings.
The irrelevant part of the two-point vertex is
 \begin{eqnarray}
 \tilde{\Gamma}_l^{(2 - \mu, Z, v)} ( Q ) & = &
 \tilde{\Gamma}_l^{(2)} ( Q ) -    
 \tilde{\Gamma}^{(2 ) }_l ( \alpha , 0 , i 0 ) 
 \nonumber
 \\
& & \hspace{-22mm}{}
-  i \epsilon 
 \left. \frac{ \partial \tilde{\Gamma}^{(2)}_l ( \alpha , 0 , i \epsilon ) }{ \partial i \epsilon } 
 \right|_{\epsilon = 0}
 - q  \left. \frac{ \partial \tilde{\Gamma}^{(2)}_l ( \alpha , q , i 0 ) }{ \partial q } 
 \right|_{q = 0}  
 \nonumber
 \\
 &  &  \hspace{-22mm} =
 \tilde{\Gamma}_l^{(2)} ( Q ) -    
 \tilde{\mu}_l -  i \epsilon ( 1 - Z_l )
 - q  ( Z_l - \tilde{v}_l )  
 \; ,
 \end{eqnarray}
where the superscripts indicate the marginal and relevant couplings to be subtracted.
The rescaled inverse propagator $r_l (q , i \epsilon )$ defined in Eq.\ (\ref{eq:rtdef})
can then be written as
  \begin{equation}
 r_l ( q , i \epsilon ) = i \epsilon - \tilde{v}_l q + \tilde{\mu}_l +
 \tilde{\Gamma}_l^{(2 - \mu, Z, v)} ( Q )
 \; .
 \label{eq:rlsub}
 \end{equation}
Finally, the irrelevant part of the four-point vertex is
 \begin{eqnarray}
  \tilde{\Gamma}_{l}^{ (4-g)}
 ( Q_1^{\prime} , {Q}_2^{\prime};  Q_2 , Q_1 )
 &  &
 \nonumber
 \\
 &  & \hspace{-35mm} =
 \tilde{\Gamma}_{l}^{ (4)}  ( Q_1^{\prime} , {Q}_2^{\prime};  Q_2 , Q_1 ) 
 - \left.
 \tilde{\Gamma}_{l}^{ (4)}  ( Q_1^{\prime} , {Q}_2^{\prime};  Q_2 , Q_1 ) 
 \right|_{ q_i = \epsilon_i =0}
 \nonumber
 \\
 &  & \hspace{-35mm} = 
 \tilde{\Gamma}_{l}^{ (4)}  ( Q_1^{\prime} , {Q}_2^{\prime};  Q_2 , Q_1 ) 
 - A_{ \alpha_1^{\prime}  \alpha_2^{\prime} ;
  \alpha_2  \alpha_1 }  \tilde{g}_l 
 \; .
 \label{eq:fourpointmom}
 \end{eqnarray}
The exact flow equations for the subtracted vertices are formally
identical with the flow equations for the un-subtracted vertices
given in Eqs.\ (\ref{eq:twopoint}) and (\ref{eq:fourpoint}), except for the fact that the
inhomogeneities 
  $ \dot{\Gamma}_l^{(2)} (Q )$ and
$ \dot{\Gamma}_{l}^{(4) }
  ( Q_1^{\prime}  , Q_2^{\prime}; 
  Q_2 , Q_1 )$ on the right-hand sides
should be replaced by their subtracted versions,
  \begin{eqnarray}
 \dot{\Gamma}_l^{(2 - \mu, Z, v)} ( Q ) & = &
 \dot{\Gamma}_l^{(2)} ( Q ) -    
 \dot{\Gamma}^{(2 ) }_l ( \alpha , 0 , i 0 ) 
 \nonumber
 \\
& & \hspace{-28mm} {}
-  i \epsilon 
 \left. \frac{ \partial \dot{\Gamma}^{(2)}_l ( \alpha , 0 , i \epsilon ) }{ \partial i \epsilon } 
 \right|_{\epsilon = 0}
 - q  \left. \frac{ \partial \dot{\Gamma}^{(2)}_l ( \alpha , q , i 0 ) }{ \partial q } 
 \right|_{q = 0}  
 \label{eq:gamma2dotsub}
 \; , \qquad
 \end{eqnarray}
 \begin{eqnarray}
  \dot{\Gamma}_{l}^{ (4-g)}
 ( Q_1^{\prime} , {Q}_2^{\prime};  Q_2 , Q_1 )
 & =  &  \dot{\Gamma}_{l}^{ (4)}  ( Q_1^{\prime} , {Q}_2^{\prime};  Q_2 , Q_1 ) 
 \nonumber
 \\
 &  & \hspace{-25mm} -
  \left.
 \dot{\Gamma}_{l}^{ (4)}  ( Q_1^{\prime} , {Q}_2^{\prime};  Q_2 , Q_1 ) 
 \right|_{ q_i = \epsilon_i =0}
 \label{eq:dotgamma4sub}
 \; . \qquad
\end{eqnarray}

\section{Approximate solution of the functional flow equations for the two-point vertex}
\setcounter{equation}{0}

So far no approximation has been made, except for the linearization of the energy dispersion.
In order to make progress, we need to truncate the hierarchy of flow equations.
We now give a truncation scheme which at weak coupling
reproduces the known scaling properties of the spectral function 
$A ( \pm ( k_F + p) , \omega )$ of the TLM. Moreover, at least
for $p = 0$, we recover the exact weak coupling result for the spectral function known from
bosonization.

\subsection{One-loop flow}

\begin{figure}[tb]
\epsfysize5.0cm 
\epsfbox{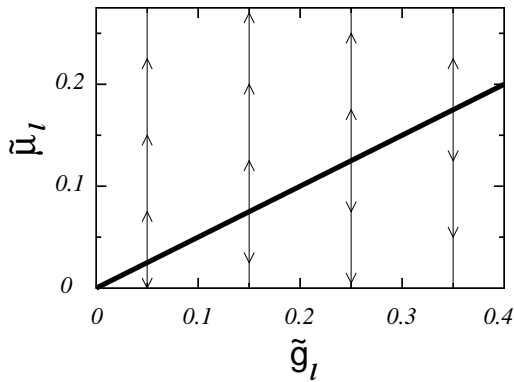}
\caption{
One-loop RG flow in the $\tilde g-\tilde{\mu}$-plane, see
Eqs.\ (\ref{eq:muoneloop}) and (\ref{eq:goneloop}).
The thick solid line is a line of  fixed points.
}
\label{fig:gmu}
\end{figure}

To begin with, let us briefly discuss the flow of the
relevant and the marginal couplings within the one-loop approximation.
We first consider the two-point vertex.
The exact flow equation  is given in
Eq.\ (\ref{eq:twopoint}). The flow is coupled to  
the irreducible four-point vertex via the inhomogeneity
$\dot{\Gamma}^{(2)}_l ( Q )$ on the right-hand side, as given in
Eq.\ (\ref{eq:B2def}).
Within the one-loop approximation, it is sufficient
to approximate the  four-point vertex appearing in
Eq.\ (\ref{eq:B2def}) by
 \begin{equation}
  \tilde{\Gamma}_{l}^{(4)}
 ( Q  , Q^{\prime} ; Q^{\prime} , Q)
 = \delta_{ \alpha , - \alpha^{\prime} } \tilde{g}_l + {\mathcal O} ( \tilde{g}_l^2 )
 \; ,
 \label{eq:gamma41loop}
 \end{equation}
where we have used the fact that $A_{ \alpha \alpha^{\prime}; \alpha^{\prime} \alpha} =
 1 - \delta_{\alpha , \alpha^{\prime} } = \delta_{\alpha , - \alpha^{\prime} }$. 
In this approximation
 \begin{eqnarray}
 \dot{\Gamma}^{(2)}_l ( Q )
 & = &  - \frac{\tilde{g}_l}{2} \sum_{\alpha^{\prime}} \int d q^{\prime} \int \frac{ d \epsilon^{\prime}}{2 \pi} \, \delta_{\alpha,-\alpha^{\prime}}
 \dot{G}_l ( Q^{\prime} ) + {\mathcal O} ( \tilde{g}_l^2 )
 \nonumber
 \\
 & = & - \frac{\tilde{g}_l}{2} \left[ \Theta ( \tilde{\mu}_l + \tilde{v}_l ) +
  \Theta ( \tilde{\mu}_l - \tilde{v}_l ) \right] + {\mathcal O} ( \tilde{g}_l^2 )
 \; . \qquad
 \end{eqnarray}
For $ | \tilde{\mu}_l | < | \tilde{v}_l |$ this simplifies to
 \begin{equation}
 \dot{\Gamma}^{(2)}_l ( Q )
  =   - \frac{\tilde{g}_l}{2} + {\mathcal O} ( \tilde{g}_l^2 )
 \; \; , \; \;  | \tilde{\mu}_l | < | \tilde{v}_l |
 \; \; .
 \end{equation}
As shown below, 
in the physically relevant case 
$\tilde{\mu}_l = {\mathcal O} ( \tilde{g}_l )$, while  $\tilde{v}_l$ is of the order of unity.
From Eqs.\ (\ref{eq:muflow}) and (\ref{eq:Ztflow})-(\ref{eq:etatB})
we conclude  that for small $\tilde{g}_l$
 \begin{eqnarray}
 \partial_l \tilde{\mu}_l & = & \tilde{\mu}_l - \frac{\tilde{g}_l}{2 }
 \; ,
 \label{eq:muoneloop}
 \\
  \partial_l Z_l & = & 0
 \; ,
 \label{eq:Zoneloop}
 \\
  \partial_l \tilde{v}_l & = & 0
 \label{eq:voneloop}
 \; .
 \end{eqnarray}
Finally, let us consider the momentum- and frequency-independent
part of the four-point vertex. 
Using the fact that $\tilde{\Gamma}^{(6)}_l$ is of order $\tilde{g}_l^3$ and hence can be 
neglected,\cite{Polchinski84,Kopietz01} we obtain from Eq.\ (\ref{eq:B4def})
to second order in $\tilde{g}_l$,
 \begin{eqnarray}
 \dot{\Gamma}_l^{(4)}
 ( {Q}_1^{\prime} , Q_2^{\prime}; 
  Q_2 , Q_1 ) & \approx &
 -  
 \frac{\tilde{g}_l^2}{2} \sum_{\alpha} \int dq \int \frac{ d \epsilon}{ 2 \pi } 
 \nonumber
 \\
 &  & \hspace{-35mm} \times \Bigl\{ \frac{1}{2}
 \bigl[ A_{ \alpha_1^{\prime}   \alpha_2^{\prime} ; \alpha^{\prime}  \alpha }
 A_{ \alpha  \alpha^{\prime} ; \alpha_2  \alpha_1 }
 \nonumber
 \\
 & & \hspace{-30mm}
 \times  \left(
\dot{G}_l ( Q  )  \tilde{G}_l 
( Q^{\prime} )  + 
 \tilde{G}_l ( {Q} )  \dot{G}_l 
 ( {Q}^{\prime} ) \right)
 \bigr]_{ K^{\prime} = K_1 + K_2 - K}
 \nonumber
 \\
 &  & 
 \hspace{-32mm} {} -  \bigl[
 A_{ \alpha_1^{\prime}   \alpha^{\prime} ; \alpha  \alpha_1 }
 A_{ \alpha_2^{\prime}  \alpha ; \alpha^{\prime}  \alpha_2 }
  \nonumber
 \\
 & & \hspace{-30mm}
\times \left(
\dot{G}_l ( Q  )  \tilde{G}_l 
( {Q}^{\prime} )  + 
 \tilde{G}_l ( Q )  \dot{G}_l 
 ( {Q}^{\prime} ) \right)
 \bigr]_{ K^{\prime} = K + K_1 - K_1^{\prime}}
 \nonumber
 \\
 &  &
 \hspace{-32mm} {}
 + 
\bigl[
 A_{ \alpha_2^{\prime}   \alpha^{\prime} ; \alpha  \alpha_1 }
 A_{ \alpha_1^{\prime}  \alpha ; \alpha^{\prime}  \alpha_2 } 
  \nonumber
 \\
 & & \hspace{-30mm}
\times
\left(
\dot{G}_l ( {Q}  )  \tilde{G}_l 
( {Q}^{\prime} )  + 
 \tilde{G}_l ( Q )  \dot{G}_l 
 ( {Q}^{\prime} ) \right)
 \bigr]_{ K^{\prime} = K + K_1 - K_2^{\prime}}
 \Bigr\} \; . 
 \nonumber
 \\
 & &
 \label{eq:F4def}
 \end{eqnarray}
Because on the critical manifold $\tilde{\mu}_l = {\mathcal O} ( \tilde{g}_l )$, 
we may approximate on the right-hand side of Eq.\ (\ref{eq:F4def})
 \begin{equation}
  \tilde{G}_l ( {Q} )  \approx \frac{ \Theta ( e^l > | q | > 1 )}{ i \epsilon - 
 \tilde{v}_l q }
  \; \; , \; \; 
 \dot{G}_l ( Q )  \approx \frac{ \delta (  | q | - 1 )}{ i \epsilon - 
 \tilde{v}_l q }
 \; .
 \label{eq:Gtapprox}
 \end{equation}
The integrations in Eq.\ (\ref{eq:F4def}) can then be performed. We obtain 
for the final result
  \begin{eqnarray}
 \dot{\Gamma}_l^{(4)}
 ( {Q}_1^{\prime} , Q_2^{\prime}; 
  Q_2 ,  Q_1 ) & \approx &
 \nonumber
 \\
 & & \hspace{-35mm}
- \tilde{g}_l^2 \Bigl\{ 
  A_{ \alpha_1^{\prime}  \alpha_2^{\prime} ; \alpha_2 \alpha_1 }
 \dot{\chi}_l ( q_1 - q_2 , i \epsilon_1 + i \epsilon_2 )
 \nonumber 
 \\
 &  & \hspace{-35mm} {} +
 E_{ \alpha_1^{\prime}  \alpha_2^{\prime} ; \alpha_2 \alpha_1 }
  \bigl[  \delta_{ \alpha_1  , \alpha_2 } 
 \dot{\Pi}_l ( q_1 - q_1^{\prime} ,  i \epsilon_1 - i \epsilon_1^{\prime} )
 \nonumber
 \\
 & &  \hspace{-19mm} {} +
 \delta_{ \alpha_1  , - \alpha_2 } 
 \dot{\chi}_l ( q_1 + q_1^{\prime} ,  i \epsilon_1 - i \epsilon_1^{\prime} )
 \bigr]
 \nonumber 
 \\
 &  & \hspace{-35mm} {} - 
 D_{ \alpha_1^{\prime}  \alpha_2^{\prime} ; \alpha_2 \alpha_1 }
  \bigl[  \delta_{ \alpha_1  , \alpha_2 } 
 \dot{\Pi}_l ( q_1 - q_2^{\prime} ,  i \epsilon_1 - i \epsilon_2^{\prime} )
 \nonumber
  \\
 & & \hspace{-19mm} {} +
 \delta_{ \alpha_1  , - \alpha_2 } 
 \dot{\chi}_l ( q_1 + q_2^{\prime} ,  i \epsilon_1 - i \epsilon_2^{\prime} )
 \bigr]
 \Bigr\}
 \label{eq:B4res}
 \; . \qquad
 \end{eqnarray}
Here
 \begin{equation}
 \dot{\chi}_l ( q , i \epsilon ) = \Theta ( e^l -1 > | q | ) 
 \frac{ \tilde{v}_l ( 2 + | q | ) }{ \epsilon^2  + \tilde{v}_l^2 ( 2 + | q | )^2 }
 \; ,
 \label{eq:chidef}
 \end{equation}
  \begin{equation}
 \dot{\Pi}_l ( q , i \epsilon ) = \Theta ( e^l + 1 > | q | > 2 ) 
 \frac{ s_q  }{ i \epsilon  + \tilde{v}_l q }
 \; ,
 \label{eq:pidef}
 \end{equation}
where we have introduced the notation
 \begin{equation}
 s_q = {\rm sign} ( q )
 \; .
 \end{equation}
Setting external momenta and frequencies equal to zero we obtain
 from Eq.\ (\ref{eq:B4res})
 \begin{eqnarray}
 \dot{\Gamma}_l^{(4)}
 ( \alpha_1^{\prime} , 0 , \alpha_2^{\prime} , 0; 
  \alpha_2 , 0 , \alpha_1, 0 )  & = & 
 - \tilde{g}_l^2   A_{ \alpha_1^{\prime}  \alpha_2^{\prime} ; \alpha_2 \alpha_1 }
 \nonumber
 \\
 & & \hspace{-40mm}  \times
\Bigl\{  
  \delta_{ \alpha_1 , \alpha_2 } 
 \dot{\chi}_l ( 0, i 0 )
 - \delta_{ \alpha_1 , \alpha_2 } \dot{\Pi}_l ( 0, i 0 )
 \Bigr\} = 0
  \; . \qquad
 \label{eq:oneloopbeta}
 \end{eqnarray}
Hence, the coefficient of order $\tilde{g}_l^2$ in the weak coupling expansion of the
function $B_l$ defined in Eq.\ (\ref{eq:Btdef}) vanishes. 
Anticipating that $\eta_l = {\mathcal O} ( \tilde{g}_l^2 )$, we obtain from 
Eq.\  (\ref{eq:gtflow}) within  the one-loop approximation
\begin{equation}
 \partial_l \tilde{g}_l  = 0
 \; .
 \label{eq:goneloop}
 \end{equation}
The vanishing of the $\beta$-function of the marginal coupling
$\tilde{g}_l$
of the TLM is well-known.
However, the one-loop RG flow of the
relevant coupling $\tilde{\mu}_l$ is non-trivial,
because the initial value of $\tilde{\mu}_l$ has to be fine-tuned
such that $\tilde{\mu}_l$ does not exhibit a runaway flow.
As emphasized
 in Ref.\ \onlinecite{Kopietz01b}, the requirement that $\tilde{\mu}_l$ flows into a 
fixed point is equivalent with the statement that
the initial $k_F$ is the correct Fermi momentum, in agreement with 
the Luttinger theorem.\cite{Luttinger60,Blagoev97}
The flow in the $\tilde{g}-\tilde{\mu}$-plane 
implied by Eqs.\ (\ref{eq:muoneloop}) and (\ref{eq:goneloop})
is shown in Fig.\  \ref{fig:gmu}.
Obviously, at the one-loop level
the marginal couplings $\tilde{g}_l$, $Z_l$ and $\tilde{v}_l$ do not flow, while
the relevant coupling $\tilde{\mu}_l$ exhibits 
a runaway flow unless
the initial value is chosen such that 
 \begin{equation}
 \tilde{\mu}_0 = \frac{\tilde{g}_0}{2}
 \; .
 \label{eq:mufinetune}
 \end{equation}
In this case $\tilde{\mu}_l = \frac{\tilde{g}_0}{2}$ for all $l$, so that we obtain a RG fixed point.
The one-loop flow equations given above have also been
discussed 
 by Shankar,\cite{Shankar94} who derived
these equations within the framework of the conventional
momentum shell technique. 
Because within the one-loop approximation
$Z_l = 1 $, this approximation is 
not sufficient  to detect the non-Fermi liquid 
behavior of our model.

\subsection{Two-loop flow of the two-point vertex}

It is now straightforward to calculate the full two-point vertex
within the two-loop approximation and to study the emergence of Luttinger liquid behavior
when the infrared cutoff is reduced.
The crucial observation is that for large values of the flow parameter $l$ the
momentum- and frequency-dependent part of the four-point vertex
$ \tilde{\Gamma}_{l}^{ (4-g)}
  ( Q_1^{\prime} , Q_2^{\prime}; 
 Q_2 , Q_1 )$  
can be expanded
in powers of the marginal coupling $\tilde{g}_l$.
This follows from the fact that
the RG trajectory approaches the manifold defined  by the relevant and marginal
couplings, so that the irrelevant couplings become
local functions of the relevant and marginal ones.\cite{Polchinski84,Kopietz01}

\begin{figure}[b]
\epsfysize5.0cm 
\epsfbox{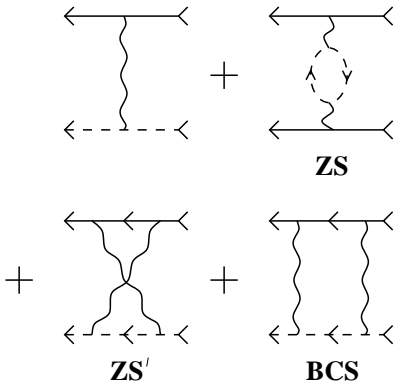}
\caption{
Feynman diagrams contributing to the effective interaction within the 
one-loop approximation, see Eq.\ (\ref{eq:veff}). 
The solid (dashed) arrows represent the fermion propagators with momenta close to the
left (right) Fermi point, and the wavy lines represent the flowing coupling
constant $\tilde{g}_l$.
}
\label{fig:Feynman}
\end{figure}

To calculate the two-point vertex at the two-loop order,
we need to know the momentum- and frequency-dependent part of the four-point vertex within the 
one-loop approximation. By definition,
\begin{widetext}
\begin{eqnarray}
 \tilde{\Gamma}_{l}^{ (4)}
 ( Q_1^{\prime} , Q_2^{\prime}; 
 Q_2 , Q_1 ) = 
  A_{ \alpha_1^{\prime}  \alpha_2^{\prime} ;
  \alpha_2  \alpha_1 } 
 \tilde{g}_l +
 \tilde{\Gamma}_{l}^{ (4-g)}
  ( Q_1^{\prime} , Q_2^{\prime};  Q_2 , Q_1 ) 
 \; ,
 \label{eq:fourpointexp}
 \end{eqnarray}
where to order $\tilde{g}_l^2$ we obtain from 
Eqs.\ (\ref{eq:Gamma4integral}) and (\ref{eq:B4res}),
 \begin{eqnarray}
  \tilde{\Gamma}_{l}^{ (4-g)}
  ( Q_1^{\prime} , Q_2^{\prime};  Q_2 , Q_1 ) 
 & \approx &
  \tilde{\Gamma}_{l=0}^{ (4-g)}
  ( Q_1^{\prime} , Q_2^{\prime};  Q_2 , Q_1 ) 
 - \tilde{g}_l^2 \Bigl\{ 
  A_{ \alpha_1^{\prime}  \alpha_2^{\prime} ; \alpha_2 \alpha_1 }
 {\chi}_{l}  ( q_1 - q_2 , i \epsilon_1 + i \epsilon_2 )
 \nonumber 
 \\
 &  & \hspace{-34mm} {} + 
 E_{ \alpha_1^{\prime}  \alpha_2^{\prime} ; \alpha_2 \alpha_1 }
  \bigl[   \delta_{ \alpha_1  , \alpha_2 } 
 {\Pi}_{l} ( q_1 - q_1^{\prime} ,  i \epsilon_1 - i \epsilon_1^{\prime} )
 +
 \delta_{ \alpha_1  , - \alpha_2 } 
 {\chi}_{l} ( q_1 + q_1^{\prime} ,  i \epsilon_1 - i \epsilon_1^{\prime} )
 \bigr]
 \nonumber 
 \\
 &  & \hspace{-34mm} {} - 
 D_{ \alpha_1^{\prime}  \alpha_2^{\prime} ; \alpha_2 \alpha_1 }
  \bigl[  \delta_{ \alpha_1  , \alpha_2 } 
 {\Pi}_{l} ( q_1 - q_2^{\prime} ,  i \epsilon_1 - i \epsilon_2^{\prime} )
  +
 \delta_{ \alpha_1  , - \alpha_2 } 
 {\chi}_{l} ( q_1 + q_2^{\prime} ,  i \epsilon_1 - i \epsilon_2^{\prime} )
 \bigr]
 \Bigr\}
 \label{eq:Gamma4momres}
 \; .
 \end{eqnarray}
Here
 \begin{equation}
  {\Pi}_{l} ( q ,  i \epsilon ) = \Pi_{l, 0} ( q , i \epsilon ) 
 \; \; , \; \; 
  {\chi}_{l} ( q ,  i \epsilon ) = \chi_{l, 0} ( q , i \epsilon ) 
 \; ,
 \end{equation}
where for arbitrary initial value $l_0 \geq 0$,
 \begin{eqnarray}
 {\chi}_{l, l_0} ( q ,  i \epsilon )  & = & \int_{l_0 }^{l}
 d t \left[ \dot{\chi}_{ l - t } ( e^{- t } q ,  e^{- t } i \epsilon )
 - \dot{\chi}_{l - t } ( 0 , i 0 ) \right]
 \nonumber
 \\
 & \approx & - \frac{1}{ 2 \tilde{v}_l }
 \biggl\{  ( l - l_0 ) \Theta ( | q| > e^l - e^{l_0} )
 +
 \frac{1}{2} \ln \left[
 \frac{ \tilde{v}_l^2 ( 2 + e^{- l_0} | q| )^2 + e^{-2 l_0} \epsilon^2 }{
 \tilde{v}_l^2 ( 2 - e^{ - l } | q| )^2 + e^{- 2 l } \epsilon^2 } \right] 
\Theta ( e^l - e^{l_0} > |q| ) 
 \biggr\}
 \; ,
\label{eq:chires}
\end{eqnarray}
and
 \begin{eqnarray}
 {\Pi}_{l, l_0} ( q ,  i \epsilon )  & = & \int_{l_0 }^{l}
 d t \left[ \dot{\Pi}_{ l - t } ( e^{- t} q ,  e^{- t } i \epsilon )
 - \dot{\Pi}_{l - t } ( 0 , i 0 ) \right]
 \nonumber
 \\
 & \approx & \frac{1}{ 2  } \; 
 \frac{ s_q}{ i \epsilon + \tilde{v}_l q } 
 \biggl\{   ( | q | - 2e^{l_0} ) \Theta (  e^l + e^{l_0} > | q| > 2 e^{l_0} )
 + ( 2 e^l - | q| ) \Theta ( 2 e^l  > |q| > e^l + e^{l_0} ) 
 \biggr\}
 \; .
\label{eq:Pires}
\end{eqnarray}
\end{widetext}
Note that Eq.\ (\ref{eq:Gamma4momres}) is manifestly antisymmetric with respect
to the exchange of the incoming or the outgoing labels.
For the calculation of the two-point vertex we only need
the irreducible four-point vertex at
$Q_1 = Q_1^{\prime}$  and $Q_2 = Q_2^{\prime}$, i.e.
 \begin{eqnarray}
  \tilde{\Gamma}_{l}^{ (4)}
 ( Q , Q^{\prime} ; Q^{\prime} , Q ) 
 & =  & \delta_{\alpha , - \alpha^{\prime} } \tilde{g}_l
 + \delta_{\alpha ,  \alpha^{\prime} } 
   \tilde{g}_l^2  \Pi_{l} ( q - q^{\prime}  , i \epsilon - i \epsilon^{\prime} )
 \nonumber
 \\
 &  & \hspace{-27mm} + \delta_{\alpha , - \alpha^{\prime} } 
   \tilde{g}_l^2 \biggl[ \chi_{l} ( q + q^{\prime} , i \epsilon - i \epsilon^{\prime} ) 
 +  \chi_{l} ( q - q^{\prime} , i \epsilon + i \epsilon^{\prime} ) 
   \biggr]
 \; .
 \nonumber
 \\
 \label{eq:veff}
 \end{eqnarray}
A graphical representation of this equation is shown
in Fig.\ \ref{fig:Feynman}.
We recognize the usual perturbative contributions to the effective
interaction: 
the first term of order $\tilde{g}_l^2$
on the right-hand side of Eq.\ (\ref{eq:veff}) corresponds to the
contribution from the zero-sound (ZS) channel, the second term corresponds to the
Peierls channel (sometimes also denoted by ZS$^\prime$ channel), 
and  the last term corresponds to the BCS channel.
Note that the zero-sound channel gives rise to an effective retarded forward scattering
interaction of the $g_4$-type (using the $g$-ology language), which is generated
by integrating out degrees of freedom. 

\begin{figure}
\epsfysize5.0cm 
\epsfbox{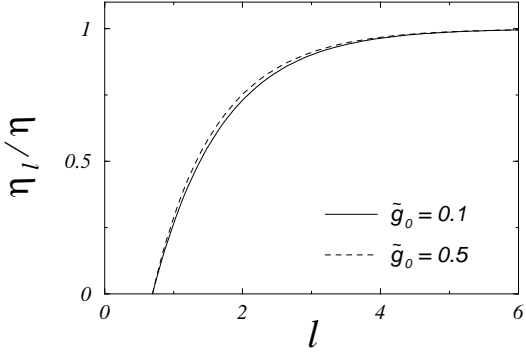}
\caption{
Anomalous dimension $\eta_l$ as a function of the logarithmic flow parameter 
$l$, see Eqs.\ (\ref{eq:etatflow}) and (\ref{eq:vflow}).
}
\label{fig:etaflow}
\end{figure}

Given an approximate expression for the four-point vertex, we may
calculate the inhomogeneity $\dot{\Gamma}_{l}^{(2) } ( Q )$
on the right-hand side of the flow equation for the two-point vertex, see
Eqs.\ (\ref{eq:twopoint}) and (\ref{eq:B2def}).
Separating the contributions due to the zero sound (ZS) and the
Peierls-BCS (PB) channels,
we have
 \begin{equation}
 \dot{\Gamma}_{l}^{(2) }
  (q , i \epsilon ) = - \frac{\tilde{g}_l}{2} +
 \dot{\Gamma}_{l}^{(2,  {\rm ZS} ) }
  ( q , i \epsilon  ) +  \dot{\Gamma}_{l}^{(2, {\rm PB}) }
  ( q , i \epsilon  ) +{\mathcal O} (\tilde{g}_l^3) 
 \; ,
 \end{equation}
where
 \begin{eqnarray}
 \dot{\Gamma}_{l}^{(2, {\rm ZS}) }
  ( q , i \epsilon  ) &  &
 \nonumber
 \\
 & & \hspace{-18mm} =
 - \frac{ \tilde{g}_l^2}{2}  \int d q^{\prime} \int \frac{ d \epsilon^{\prime} }{ 2 \pi}  
  \frac{ \delta ( | q^{\prime} | -1 )}{ i \epsilon^{\prime} - \tilde{v}_l q^{\prime} } \,
  \Pi_{l} ( q - q^{\prime} , i \epsilon - i \epsilon^{\prime} )
 \nonumber
 \\
 &  & \hspace{-18mm} = 
 - \frac{ \tilde{g}_l^2}{4} s_q
 \Bigl[ 
\Theta ( e^l > | q | > 1 ) \,
 \frac{ | q | -1}{ \tilde{v}_l  ( 2 + | q | ) + 
  i \epsilon s_q  }  
 \nonumber
 \\
 & & \hspace{-14mm} {}
 + \Theta ( 2 e^l -1 > | q | > e^l) \,
 \frac{ 2 e^l -1 - | q | }{ \tilde{v}_l  ( 2 + | q | ) + 
  i \epsilon s_q  }  
 \Bigr]
 \; , \qquad \quad
 \label{eq:Gamma2dotpi}
 \end{eqnarray}
 \begin{eqnarray}
 \dot{\Gamma}_{l}^{(2, {\rm PB}) }
  ( q , i \epsilon  ) 
 & = &
 - \frac{ \tilde{g}_l^2}{2}  \int d q^{\prime} \int \frac{ d \epsilon^{\prime} }{ 2 \pi}  
  \frac{ \delta ( | q^{\prime} | -1 )}{ i \epsilon^{\prime} - \tilde{v}_l q^{\prime} }
 \nonumber
 \\
 & & \hspace{-15mm} \times  
 \left[ \chi_{l} ( q + q^{\prime} , i \epsilon - i \epsilon^{\prime} )
 -  \chi_{l} ( q - q^{\prime} , i \epsilon + i \epsilon^{\prime} )
 \right]
 \nonumber
 \\
&  & \hspace{-25mm} = 
 \frac{\tilde{g}_l^2}{ 4 \tilde{v}_l } s_q  \left\{
 \Theta ( 1 > | q | > 2 - e^l    ) 
 \ln \left[ \frac{ \displaystyle  \tilde{v}_l ( 4  - | q | ) + i \epsilon s_q }{ 
 \displaystyle \tilde{v}_l ( 2 e^l + | q| ) + i \epsilon s_q } \right]
 \right.
 \nonumber
 \\
 & & 
 \hspace{-21.5mm} {}
 +
 \Theta (   e^l >  | q | > 1  )
 \ln \left[ \frac{ \displaystyle  \tilde{v}_l ( 2  + | q | ) + i \epsilon s_q }{ 
 \displaystyle \tilde{v}_l ( 2 e^l + 2 - | q| ) + i \epsilon s_q } \right]
 \nonumber
 \\
 & & \left. \hspace{-22mm} {} - 
 \Theta (   e^l -2 > | q | ) 
 \ln \left[ \frac{ \displaystyle  \tilde{v}_l ( 4  + | q | ) - i \epsilon s_q }{ 
 \displaystyle \tilde{v}_l ( 2 e^l - | q| ) - i \epsilon s_q } \right]
 \right\}
 \; .
 \label{eq:Gammadotchi3}
\end{eqnarray}
Expanding to first order in $ q$ and $ \epsilon$, we obtain
 \begin{eqnarray}
 \dot{\Gamma}_{l}^{(2, {\rm PB}) }
  ( q , i \epsilon  ) & = & 
 \frac{\tilde{g}_l^2}{ 8 \tilde{v}_l^2 }  \Theta (  e^l -2 ) 
 \Bigl[ (1 -2 e^{-l} ) i \epsilon 
 \nonumber
 \\
 \!\!\!\!\!\!\!\! & & 
- ( 1 + 2 e^{-l}) \tilde{v}_l q  
 + {\mathcal O} ( \epsilon^2 , q^2 , \epsilon q ) \Bigr]
 \; . \qquad
 \end{eqnarray}  
Because $ \dot{\Gamma}_{l}^{(2, {\rm ZS} ) }
  ( q , i \epsilon  )$ vanishes for $ | q | < 1$, it does not contribute
to the flow of the marginal couplings $Z_l$ and $\tilde{v}_l$.
From Eq.\ (\ref{eq:etatB}) we obtain for the anomalous dimension
at scale $l$,
 \begin{equation}
 \eta_l = \frac{\tilde{g}_l^2}{8 \tilde{v}_l^2 } \Theta ( e^l -2 ) ( 1 - 2 e^{-l} )
 \label{eq:etatflow}
 \; ,
 \end{equation}
and from Eq.\ (\ref{eq:vtflow}) we find for the flow of the 
Fermi velocity renormalization factor
 \begin{equation}
 \partial_l \tilde{v}_l = 
  \frac{\tilde{g}_l^2}{2 \tilde{v}_l } \Theta ( e^l -2 ) e^{-l}
 \; .
 \label{eq:vflow}
 \end{equation}
Using the fact that $\tilde{g}_l = \tilde{g}$ is independent of 
the flow parameter $l$,
Eq.\ (\ref{eq:vflow}) can be easily integrated,
 \begin{equation}
 \tilde{v}_l = \tilde{v}_0 \Bigl[ 1 +  
 \Theta ( e^l -  2 ) \frac{\tilde{g}^2}{2 \tilde{v}_0^2 } ( 1 - 2 e^{-l } ) 
 \Bigr]^{1/2}
 \; .
 \end{equation} 
Note that $\tilde{v}_l$ rapidly approaches a constant
$\tilde{v}_{\infty}$ for large $l$, which is to leading order in $\tilde g$ given by 
 \begin{equation}
 \tilde{v}_{\infty}  = \tilde{v}_0 + 
 \frac{\tilde{g}^2}{4 \tilde{v}_0 }  + {\mathcal O} ( \tilde{g}^4 )
 \; .
 \label{eq:tildevfix}
 \end{equation}
In Fig.\ \ref{fig:etaflow} we show
the flow of the  anomalous dimension $ \eta_l$ given in 
Eq.\ (\ref{eq:etatflow}) as a function
of the logarithmic flow parameter $l$.
Obviously, $\eta_l$ vanishes for
$ l < \ln 2$, and approaches a constant $\eta$ for large $l$, which is given by
 \begin{equation}
 \eta = \frac{\tilde{g}^2}{8 \tilde{v}_{\infty} }
 \label{eq:etafix}
 \; .
 \end{equation}
The above fixed point value of $\eta$ agrees with the
weak coupling expansion of the bosonization result 
for the anomalous dimension of the spinless $g_2$-TLM.
Note, however,  that the exact  solubility of  the TLM
relies on the filled Dirac sea associated with the
linearized energy dispersion. 
As discussed by Schulz and Shastry,\cite{Schulz98}
the introduction of the Dirac sea leads to finite renormalizations of the
fixed point values of the Luttinger liquid parameters  (such as the anomalous dimension), 
so that the value of $\eta$ beyond the leading order $\tilde g$ is modified.
This can be explicitly verified within our RG approach, where
the introduction of the Dirac sea corresponds
to the removal of the ultraviolet cutoff $\Lambda_0$. Then we should 
take the limit  $e^{l} \rightarrow \infty$ in
Eqs.\ (\ref{eq:Gamma2dotpi})
and (\ref{eq:Gammadotchi3}), so that the
explicit $l$-dependence on the right-hand sides of these expressions disappears.
In this case the marginal coupling $\tilde{v}_l$  
is  not renormalized at all ($\partial_l \tilde{v}_l = 0$), so that 
$\tilde{v}_\infty = \tilde{v}_0$.
Furthermore,  the initial flow of the anomalous dimension $\eta_l$ 
discussed above is
also  removed by the Dirac sea, so that  $\eta_l$ is effectively replaced by its 
asymptotic limit $\eta$ given in Eq.\ (\ref{eq:etafix}). 
To simplify the following calculation of the spectral function, we shall 
from now on work with a Dirac sea, which amounts to
taking the limits $\Lambda_0 \rightarrow \infty$ and
$l \rightarrow \infty$ in
Eqs.\ (\ref{eq:Gamma2dotpi})
and (\ref{eq:Gammadotchi3}). Then  we obtain
\begin{eqnarray}
 \dot{\Gamma}_{\infty}^{(2, {\rm ZS}) }
  ( q , i \epsilon  ) &  = &  -  \Theta ( | q | > 1 )    2 \eta  s_q 
  \frac{ | q | -1 }{  \tilde{v}_0 ( 2 + | q |)   + i \epsilon s_q }
 \nonumber
 \\
 &  & \hspace{-20mm} = 
 \Theta ( | q | > 1 )    \eta s_q
 \left[ 1 - \frac{ 3  \tilde{v}_0 | q | + i \epsilon s_q }{  \tilde{v}_0 ( 2 + | q | )  + i
 \epsilon s_q } \right] 
 \; ,
 \label{eq:ZSdirac}
 \end{eqnarray}
\begin{eqnarray}
 \dot{\Gamma}_{\infty}^{(2, {\rm PB}) }
  ( q , i \epsilon ) = 
 2 \eta  s_q 
 & & \!\!\!\!\!\!
 \biggl\{
 \Theta (  | q | > 1 ) \ln [  \tilde{v}_0 ( 2 + | q | )  + i \epsilon s_q ]
  \nonumber
 \\
 & & \!\!\!\!\!\! {}+
  \Theta ( 1 >  | q |  ) \ln [  \tilde{v}_0 ( 4 - | q |)  + i \epsilon s_q ]
 \nonumber
 \\
 & & \!\!\!\!\!\! -  \ln [  \tilde{v}_0 ( 4 + | q |)  - i \epsilon s_q ]
 \biggr\}
 \; .
 \label{eq:PBdirac}
\end{eqnarray}  
The corresponding subtracted function $ \dot{\Gamma}_{\infty}^{(2 - \mu,Z,v) } ( Q )$
defined in Eq.\ (\ref{eq:gamma2dotsub}) is in this approximation given by
 \begin{eqnarray}
  \dot{\Gamma}_{\infty}^{(2 - \mu,Z,v ) } ( q , i \epsilon ) 
  & = &    \eta s_q 
  \biggl\{  
 2 \Theta (  | q | > 1 ) \ln \left[ \tilde{v}_0 ( 2 + | q | ) + i \epsilon s_q \right]
 \nonumber
 \\
 & &   
 \hspace{-1mm} {} +
  2 \Theta ( 1 >  | q |  ) \ln \left[ \tilde{v}_0 ( 4 - | q | ) + i \epsilon s_q \right]
 \nonumber
 \\
 & &  \hspace{-1mm} {} -  2 \ln \left[ \tilde{v}_0 ( 4 + | q | ) - i \epsilon s_q \right]
 \nonumber
 \\
 &   & \hspace{-1mm} {} +   
 \Theta ( | q | > 1 ) \left[ 1 - 
  \frac{ 3 \tilde{v}_0 | q | + i \epsilon s_q }{  \tilde{v}_0 (2 +
   | q | ) + i \epsilon s_q } \right]
 \nonumber
 \\
 & &   \hspace{-1mm} {} - ( i \epsilon s_q - \tilde{v}_{0} | q  | )
 \biggr\} 
 \; .
 \label{eq:gammadotsubfixed}
 \end{eqnarray}
So far we have calculated the function
 $ \dot{\Gamma}_{\infty}^{(2 - \mu,Z,v) } ( Q )$
defined in Eq.\ (\ref{eq:gamma2dotsub})  perturbatively to second order in $\tilde{g}$. 
Let us for the moment proceed
further within perturbation theory. In this case we may set $\eta_l = 0$
on the right-hand side of the exact flow equation (\ref{eq:twopoint}) for the two-point vertex, 
which amounts to setting $\bar{\eta}_l ( t ) = 0$ in Eq.\ (\ref{eq:twopointint}).
Assuming that initially $\tilde{\Gamma}^{(2 - \mu, Z , v )}_{l=0} ( Q) = 0$, we obtain
in this approximation
\begin{eqnarray}
  & \tilde{\Gamma}^{(2 - \mu, Z , v ) }_l  ( q , i \epsilon  )  
 \approx  \int_{0}^{l} d t \, e^t 
 \dot{\Gamma}_{\infty}^{(2 - \mu , Z , v ) } ( e^{- t} q , e^{-t} i \epsilon ) &
 \nonumber
 \\
 &  =  \int_{e^{- l}}^{1} d \lambda \, \lambda^{-2 }  
 \dot{\Gamma}_{\infty}^{(2 - \mu , Z , v ) } ( \lambda q , \lambda i \epsilon )
 \; . &
 \label{eq:gammasubintpert}
 \end{eqnarray}
Substituting the perturbative result (\ref{eq:gammadotsubfixed}) 
into Eq.\ (\ref{eq:gammasubintpert}),
going back to unrescaled variables $p = \Lambda q / v_F $,
$\omega =  \Lambda \epsilon $, 
 $ \tilde{\Gamma}_{l}^{(2)}
 ( Q  )  =
 - \frac{Z_l }{\Lambda} 
 \left[ \Sigma_{\Lambda} ( k , i \omega ) - \Sigma ( \alpha k_F , i0 ) \right]$
(see Eq.\ (\ref{eq:Gammatdef})), and finally taking the scaling limit
$l \rightarrow \infty$ (i.e. $\Lambda \rightarrow 0$),
we recover the well-known
perturbative self-energy of the spinless $g_2$-TLM,
 \begin{eqnarray}
 \Sigma  ( \alpha  k_F+  \alpha p  , i \omega ) 
 - \Sigma  ( \alpha  k_F  , i \omega ) 
 & \approx &
 \nonumber
 \\ & & \hspace{-46mm}
 - \frac{ \tilde{g}^2}{8} {v}_F p + \frac{ \tilde{g}^2}{16}
 ( i \omega - {v}_F p ) \ln \left[ \frac{ ( {v}_F p )^2 + \omega^2 }{ \xi_{0}^2 } 
 \right]
 \; . \qquad
 \label{eq:sigmapert}
 \end{eqnarray}
It is also easy to check that in the regime
 \begin{equation}
  1 \ll | q |  \ll e^{l} \; \; , \; \;  1 \ll  | \epsilon | \ll e^{l}
 \; ,
 \label{eq:scaling}
 \end{equation}
we may approximate
 \begin{equation}
  \tilde{\Gamma}^{(2 - \mu, Z , v ) }_l  ( q , i \epsilon  )   
 \approx \tilde{\Gamma}^{(2 - \mu, Z , v ) }_{l \rightarrow \infty}  ( q , i \epsilon  )   
 \; , 
 \label{eq:gammalargel}
\end{equation}
so that the only $l$-dependence of the self-energy 
enters via the scaling variables $ q = p \xi = p \xi_0 e^{l}$ and
$ \epsilon = \omega \tau = \omega  / \Lambda = \omega e^{l} / \Lambda_0$.
Thus, in  the scaling limit the spectral function can be written in terms
of a scaling function which does not explicitly depend on the 
logarithmic scale factor $l$.

We now propose a simple procedure to go beyond perturbation
theory: from Eq.\ (\ref{eq:twopointint}) we know that the
exact two-point vertex for $l \gg 1$ satisfies 
\begin{eqnarray}
  \tilde{\Gamma}^{(2 - \mu, Z , v ) }_l  ( q , i \epsilon  )  
 & =  & \int_{0}^{l} d t \, e^{ (1 - \eta ) t } \,
 \dot{\Gamma}_{\infty}^{(2 - \mu , Z , v ) } ( e^{- t} q , e^{-t} i \epsilon )
 \nonumber
 \\
 &   &  \hspace{-14mm} = \int_{e^{-l}}^{1} d \lambda \, \lambda^{-2 + \eta} \, 
 \dot{\Gamma}_{\infty}^{(2 - \mu , Z , v ) } ( \lambda q , \lambda i \epsilon )
 \; , \qquad
 \label{eq:gammasubint}
 \end{eqnarray}
which differs from the perturbative expression (\ref{eq:gammasubintpert})
because the fixed-point value of the anomalous dimension 
appears on the right-hand side. 
We now approximate the flow function
$ \dot{\Gamma}_{\infty}^{(2 - \mu , Z , v ) } (  q , i \epsilon )$
on the right-hand side of Eq.\ (\ref{eq:gammasubint}) by
 \begin{eqnarray}
  \dot{\Gamma}_{\infty}^{(2 - \mu,Z,v ) } ( q , i \epsilon ) 
  & \approx &    \eta s_q 
  \biggl\{  
 2 \Theta (  | q | > 1 ) \ln \left[ \tilde{v} ( 2 + | q | ) + i \epsilon s_q \right]
 \nonumber
 \\
 & &   
 \hspace{0mm} {} +  2 \Theta ( 1 >  | q |  ) \ln \left[ \tilde{v} ( 4 - | q | ) + i \epsilon s_q \right]
 \nonumber
 \\
 & &  \hspace{0mm} {} -  2 \ln \left[ \tilde{v} ( 4 + | q | ) - i \epsilon s_q \right]
 \nonumber
 \\
 &   &  {} -   
 \Theta ( | q | > 1 ) 
  \frac{ 3 \tilde{v} | q | + i \epsilon s_q }{  \tilde{v} (2 +
   | q | ) + i \epsilon s_q } 
 \nonumber
 \\
 & &  {} -  ( i \epsilon s_q - \tilde{v} | q  | )
 \biggr\} 
 \; .
 \label{eq:gammadotsubmod}
 \end{eqnarray}
This expression is almost identical with the 
perturbative two-loop result (\ref{eq:gammadotsubfixed}), except that
the term $\eta s_q \Theta ( | q | > 1 )$ has been omitted, and the factor $\tilde{v}_0$
has been replaced by the 
perturbative Fermi velocity  renormalization factor
 \begin{equation}
 \tilde{v} = 1  - \frac{\tilde g^2}{8} = 1 - \eta
 \; ,
 \label{eq:vren}
 \end{equation}
see Eq.\ (\ref{eq:vcbos}).
Note that within perturbation theory 
the term  $\eta s_q \Theta ( | q | > 1 )$ in Eq.\ (\ref{eq:gammadotsubfixed})
is responsible  for the finite Fermi velocity renormalization
given by the first term $ - \frac{\tilde{g}^2}{8} v_F p$ on the right-hand side of
Eq.\ (\ref{eq:sigmapert}).  By substituting $ \tilde{v}_0 \rightarrow \tilde{v}$
in Eq.\ (\ref{eq:gammadotsubfixed}), we have implicitly replaced the bare
propagator in the scaling limit by $ [ i \epsilon - \tilde{v} q ]^{-1}$,
thus taking the Fermi velocity renormalization 
in the scaling limit 
(where the dimensionless variables $p$ and $\epsilon$ are large compared with unity)
self-consistently into account. 
Note that according to  Eq.\ (\ref{eq:meffren}) 
the dimensionless quantity $\tilde{v}$ can be
identified with the  renormalization of the
inverse effective mass, which remains finite in $1d$.

\begin{figure}[t]
\epsfysize5.8cm 
\epsfbox{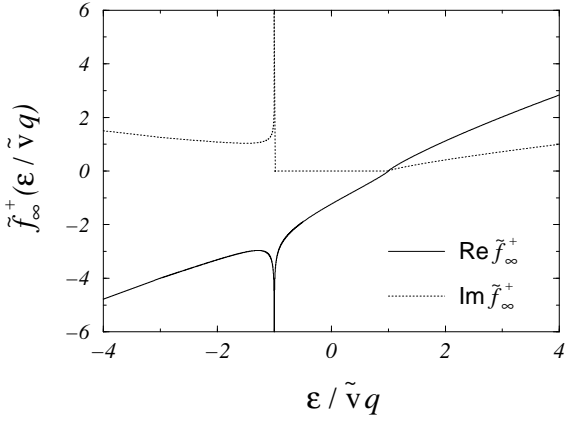}
\caption{
Real and imaginary part of $\tilde f_{\infty}^{+} (\epsilon/\tilde v q) \equiv 
r_{\infty} (q,\epsilon+i0)/|q|^{1-\eta} = \left[(\epsilon -\tilde v q) + 
\tilde{\Gamma}^{(2-\mu , Z, v ) }_\infty ( q ,  \epsilon + i 0)\right]/|q|^{1-\eta}$ 
for $\eta = 0.2$, where 
$\tilde{\Gamma}^{(2-\mu , Z, v ) }_\infty ( q ,  \epsilon + i 0)$ is 
the subtracted irreducible two-point scaling function, see
Eqs.\ (\ref{eq:Imsigmascale}) and (\ref{eq:Resigmascale}).
}
\label{fig:self}
\end{figure}

Given Eqs.\ (\ref{eq:gammasubint}) and (\ref{eq:gammadotsubmod}),
we may calculate the complete dynamic scaling function
for the irreducible self-energy, describing the change in scaling behavior
as we move from the hydrodynamic into the scaling regime.
In the hydrodynamic regime the scaling function is analytic in
$ q $ and $\epsilon$. On the other hand, for large $ | q |$ and
$ | \epsilon |$ the scaling function exhibits algebraic singularities
corresponding to Luttinger liquid behavior.
In the scaling regime (\ref{eq:scaling}) we may 
use the approximation (\ref{eq:gammalargel}) and obtain
from Eqs.\ (\ref{eq:gammasubint}) and (\ref{eq:gammadotsubmod})
\begin{eqnarray}
 \tilde{\Gamma}^{(2 - \mu,Z,v ) }_\infty ( q , i \epsilon ) & = &
\frac{\eta }{2} ( i \epsilon  - \tilde{v} q )
 \nonumber
 \\
 & & \hspace{-24mm} \times
 \Biggr\{  
 f  \left( \frac{ i \epsilon s_q - \tilde{v} | q | }{ 4 \tilde{v} }  \right)
 + f \left( - \frac{ i \epsilon s_q + \tilde{v} | q | }{ 2 \tilde{v}  }  \right)
 \nonumber
 \\
 &  &  \hspace{-24mm} {} +   | q |^{- \eta }
 \biggr[  
 f \left( - \frac{ i \epsilon s_q -  \tilde{v} | q | }{ 4 \tilde{v}  | q | }  \right)
 -
 f  \left( - \frac{ i \epsilon s_q +  \tilde{v} | q | }{ 2 \tilde{v}  | q | }  \right)
\biggl]
 \Biggl\}
 \; ,
 \nonumber
 \\
 & & 
 \label{eq:gammaresscaling}
 \end{eqnarray}
where the complex function $f ( z)$ is defined by
 \begin{equation}
 f ( z ) =  z \int_{0}^{1} d \lambda \, \frac{ \lambda^{\eta} }{  1 - z \lambda }
 = z \;   _{2}F_{1} ( 1 , 1 + \eta , 2 + \eta ; z )
 \label{eq:fdef}
 \; .
 \end{equation}
Here, $_{2}F_{1} (a,b,c;z)$ is a hypergeometric function.
Note that Eq.\ (\ref{eq:gammaresscaling}) is restricted to the scaling regime
defined by (\ref{eq:scaling}).
To obtain the spectral function, we analytically continue
 $  \tilde{\Gamma}^{(2-\mu , Z, v ) }_\infty ( q ,  i \epsilon )$
according to $ i \epsilon \rightarrow \epsilon + i 0$. Using
 \begin{equation}
 {\rm Im}\, f ( x \pm i0 ) = \pm \pi \Theta ( x > 1 ) x^{- \eta}
 \; ,
 \end{equation}
and assuming that
$  | \epsilon^2 - \tilde{v}^2 q^2 | \gg 1$, we obtain
 \begin{eqnarray}
  {\rm Im}\,    \tilde{\Gamma}^{(2-\mu , Z, v ) }_\infty ( q ,  \epsilon + i 0) & = &
 2 \pi \eta 
 \nonumber
 \\
 & & \hspace{-29mm}
 \times \Biggl\{ \Theta ( \epsilon s_q > \tilde{v} | q |  ) 
 \left| \frac{ \epsilon - \tilde{v} q }{ 4 \tilde{v} } \right|^{1 - \eta}
 \nonumber
 \\
 & & \hspace{-27mm} {} +
  \Theta ( -\tilde{v}  | q |  >  \epsilon s_q   > 
  - 3 \tilde{v}   | q | )
\left| \frac{ \epsilon - \tilde{v} q }{ 4 \tilde{v} } \right|
 \left| \frac{ \epsilon + \tilde{v} q }{ 2 \tilde{v}} \right|^{ - \eta }
 \nonumber
 \\
 & & \hspace{-27mm} {} +
 \Theta ( -3 \tilde{v}  | q |  >  \epsilon s_q    )
\left| \frac{ \epsilon - \tilde{v} q }{ 4 \tilde{v} } \right|^{ 1 - \eta}
 \Biggr\}
 \; .
 \label{eq:Imsigmascale}
 \end{eqnarray}
Mathematically the imaginary part of  
$\tilde{\Gamma}^{(2-\mu , Z, v ) }_\infty ( q ,  \epsilon + i 0)$ 
arises from the branch cut
of $ f ( x + i y  )$ on the real axis for $ x > 1$. 
Furthermore, 
using the fact that for
$ | {\rm arg} (-z ) | < \pi$,
 \begin{equation}
 f ( z ) = \frac{ \pi (1 + \eta )}{ \sin ( \pi \eta ) }
 ( - z )^{ - \eta }
 - {\frac{ 1 + \eta }{\eta } }    {_{2}F_1} ( 1 , - \eta , 1 - \eta ; 
 \frac{1}{z} )
 \; ,
 \label{eq:fasym}
 \end{equation}
we obtain in the regime $ |  \epsilon^2 - \tilde{v}^2 q^2 | \gg 1$ 
and $\eta \ll 1$ for the real part, 
  \begin{eqnarray}
  {\rm Re}\,    \tilde{\Gamma}^{(2-\mu , Z, v ) }_\infty ( q ,  \epsilon + i 0) & = &
 - ( \epsilon - \tilde{v} q )
 \nonumber
 \\
 & & \hspace{-37mm} {} +
   \Theta (   \epsilon s_q   > 
  - 3 \tilde{v}   | q |  )    ( \epsilon - \tilde{v} q ) \frac{1}{2} 
  \left[
 \left| \frac{ \epsilon - \tilde{v} q }{ 4 \tilde{v} } \right|^{  - \eta} 
 + 
 \left| \frac{ \epsilon + \tilde{v} q }{ 2 \tilde{v} } \right|^{ - \eta } \right]
 \nonumber
 \\
 & & \hspace{-37mm} {} +
 \Theta ( -3 \tilde{v}  | q |  >  \epsilon s_q    )
 ( \epsilon - \tilde{v} q )
\left| \frac{ \epsilon - \tilde{v} q }{ 4 \tilde{v} } \right|^{ - \eta}
 \; .
 \label{eq:Resigmascale}
 \end{eqnarray}
In deriving this expression, we have neglected corrections of the order of $\eta$ 
to the prefactor. Note that the imaginary part given in Eq.\ (\ref{eq:Imsigmascale}) is proportional
to $\eta$, while no such small prefactor
appears in the real part, Eq.\  (\ref{eq:Resigmascale}).   
The first term on the right-hand side of Eq.\  (\ref{eq:Resigmascale})   exactly cancels
the inverse bare propagator $ i\epsilon - \tilde{v} q $ such that 
$r_{\infty} (q,\epsilon+i0) = (\epsilon -\tilde v q) + 
\tilde{\Gamma}^{(2-\mu , Z, v ) }_\infty ( q ,  \epsilon + i 0)$
has the dynamic scaling property\cite{Halperin69}
\begin{equation}
  \label{eq:sc:r}
  r_{\infty} (q,\epsilon+i0) = |q|^{1-\eta} \tilde f_{\infty}^{\pm} (\epsilon/\tilde v q) \;,
\end{equation}
where the functions $f_{\infty}^{+}(x)$ and $f_{\infty}^{-}(x)$ refer to $q>0$ and $q<0$ respectively and $(f_{\infty}^{+}(x))^{\ast} = -f_{\infty}^{-}(x)$.
A graph of $\tilde f_{\infty}^{+} (\epsilon/\tilde v q)$ 
is shown in Fig.\ \ref{fig:self} for $\eta = 0.2$.

\begin{figure}[b]
\epsfysize5.8cm 
\hspace{5mm}
\epsfbox{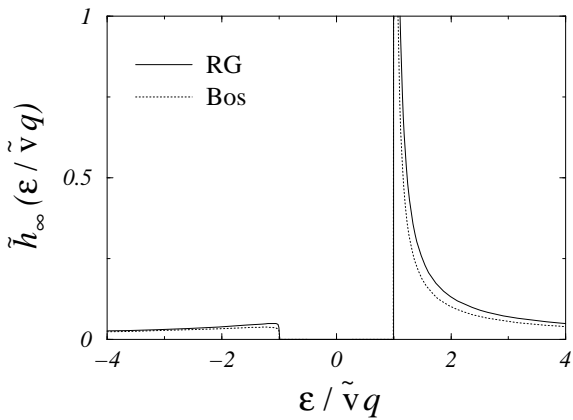}
\vspace{5mm}
\caption{Graph of the dynamic scaling function $\tilde{h}_{\infty}  (\epsilon/\tilde v q ) \equiv \tilde{A}_{\infty} (q,\epsilon) / |q|^{\eta - 1}$ 
for $ \eta =0.2$, see Eqs.\  (\ref{eq:specRGscale}) and   (\ref{eq:AtildeRG}).
For smaller $\eta$ the RG result is indistinguishable from the bosonization 
result $\tilde{h}_{\text{TL}}  (\epsilon/\tilde v q )$ 
on this scale.
}
\label{fig:spec}
\end{figure}

Using the fact that for $ \xi \gg \xi_0$ we may identify
 \begin{eqnarray}
 Z_l = Z_0 \, e^{-\int_0^l dt\, \eta_t} \equiv \bar Z_0 \, e^{- \eta l }  \approx \left( \frac{ \xi_0}{\xi} \right)^{\eta} 
 \; ,
 \end{eqnarray}
where the constant $\bar Z_0 = {\mathcal{O}}(1)$ has been absorbed by $\xi_0$,
the corresponding spectral function 
can be written in the scaling form similar to Eq.\ (\ref{eq:dynscale}),
  \begin{equation}
 A (  \alpha(k_F + p) , 
 \omega )  =  \tau  \left( \frac{\xi_0}{\xi } \right)^{  \eta } 
 \tilde{A}_{\infty} ( p \xi , \omega \tau ) 
 \;  ,
 \label{eq:specRGscale} 
\end{equation} 
where $\tau = \xi / v_F = 1 / \Lambda$, and
the scaling function $ \tilde{A}_{\infty} ( q , \epsilon ) $ is 
\begin{widetext}
 \begin{eqnarray}
 \tilde{A}_{\infty} ( q , \epsilon )  & = &
 - \frac{1}{ \pi} \, {\rm Im} \left[ \frac{1}{ \epsilon - \tilde{v} q + \tilde{\mu}_\infty + 
 \tilde{\Gamma}^{(2-\mu , Z, v ) }_\infty ( q ,  \epsilon + i 0) }
 \right]
=   \frac{\eta}{2}  
  | \epsilon - \tilde{v} q |^{ \eta -1 } 
 \nonumber
 \\
 & & \hspace{2mm}
{{ \times 
 \left\{
 \begin{array}{ll} 
 4 \left[ 1 + \left| \frac{ \epsilon - \tilde{v} q }{ \epsilon + \tilde{v} q}
 \right|^{\eta} \right]^{-2}  &  \mbox{for $ \epsilon s_q > \tilde{v} | q | $}
 \\
 0  &  \mbox{for $  \tilde{v} | q | > \epsilon s_q > - \tilde{v} | q | $}
 \\
 4 \left[ 
\left| 2 \frac{ \epsilon + \tilde{v} q }{ \epsilon - \tilde{v} q}
 \right|^{ \eta/2} 
+
 \left| \frac{1}{2} \frac{ \epsilon - \tilde{v} q }{ \epsilon + \tilde{v} q}
 \right|^{\eta/2} 
\right]^{-2 } \qquad & \mbox{for $ - \tilde{v} | q | > \epsilon s_q > - 3 \tilde{v} | q | $}
 \\
 1
 &  \mbox{for $ -3 \tilde{v} | q | > \epsilon s_q \; .$ }
 \end{array}
 \right. 
  }}
 \label{eq:AtildeRG}
 \end{eqnarray}
\end{widetext}
In the prefactor we have retained only the leading order in $\eta$.
For $q=0$, the dynamic scaling function reduces to
\begin{equation}
  \label{eq:sf:q=0}
   \tilde{A}_{\infty} (0,\epsilon) =
  \frac{\eta}{2} 
  |\epsilon|^{ \eta -1 } \;.
\end{equation}
If $q \neq 0 $, the dynamic scaling function can be written in the
dynamic scaling form\cite{Halperin69}
\begin{equation}
  \label{eq:sf:qneq0}
   \tilde{A}_{\infty} (q,\epsilon) = |q|^{\eta - 1} \tilde h_{\infty} (\epsilon/\tilde v q) \;,
\end{equation}
where $\tilde h_{\infty} (x)$ is defined via Eq.\ (\ref{eq:AtildeRG}).
A plot of the dynamic scaling function $\tilde h_{\infty}  (\epsilon/\tilde v q)$
is given in Fig.\ \ref{fig:spec} for $\eta =0.2$.
For a comparison, we have also plotted the generally accepted scaling function 
of the TLM, given in Eq.\ (\ref{eq:tildeAscale}). The qualitative agreement between the two plots becomes even better for smaller $\eta$.
Recall that Eq.\ (\ref{eq:AtildeRG}) is only valid for
$ | q | \gg 1$, $| \epsilon | \gg 1$ {\it{and for large}} 
$\big| | \epsilon| - \tilde{v} |q| \big|$.
In this limit
the small constant $\tilde{\mu}_\infty$ on the
right-hand side of Eq.\ (\ref{eq:AtildeRG})
can be neglected.
In terms of the physical variables
 $ p = q / \xi$ and $ \omega = \epsilon / \tau = \epsilon v_F / \xi$ the
condition $\big| |\epsilon| - \tilde{v} |q| \big| \gg 1$ becomes
 $\big| | \omega| - v_c |p| \big| \gg 1/ \xi$, where $v_c = \tilde{v} v_F$.
Using the fact that the function
$\tilde{A}_{\infty} ( q , \epsilon )$ satisfies the scaling relation
\begin{equation}
 \tilde{A}_{\infty} ( s q , s \epsilon ) = s^{\eta -1 }
 \tilde{A}_{\infty} ( q , \epsilon )
 \label{eq:scalingA}
\end{equation}
we may write $\tilde{A}_{\infty} ( p \xi , \omega \tau ) = \tau^{\eta -1} 
 \tilde{A}_{\infty} ( v_F p , \omega )$. With
 $\tau ( \xi_0 / \xi )^{\eta}  \tau^{\eta -1 } = \Lambda_0^{- \eta }$ we finally obtain
for the spectral function at the Luttinger liquid fixed point, 
\begin{widetext}
\begin{eqnarray}
 A (  \alpha (k_F + p)  , 
 \omega )  & = &  \Lambda_0^{- \eta} \frac{\eta}{2}
 | \omega - {v}_c p |^{ \eta -1 }
 \nonumber 
 \\
 & & \hspace{-5mm}
 {{ 
 \times
 \left\{
 \begin{array}{ll} 
 4 \left[ 1 + \left| \frac{ \omega - {v}_c p }{ \omega + {v}_c p}
 \right|^{\eta} \right]^{-2}  &  \mbox{for $ \omega s_p > {v}_c | p | $}
 \\
 0  &  \mbox{for $  {v}_c | p | > \omega s_p > - {v}_c | p | $}
 \\
 4 \left[ 
\left| 2 \frac{ \omega + {v}_c p }{ \omega - {v}_c p}
 \right|^{ \eta/2} 
+
 \left| \frac{1}{2} \frac{ \omega - {v}_c p }{ \omega + {v}_c p}
 \right|^{\eta/2} 
\right]^{-2 } \qquad & \mbox{for $ - {v}_c | p | > \omega s_p > - 3 {v}_c | p | $}
 \\
 1
 &  \mbox{for $ -3 {v}_c | p | > \omega s_p \; .$ }
 \end{array}
 \right. 
 }}
 \label{eq:ARG2}
 \end{eqnarray}
\end{widetext}
Let us now compare Eq.\ (\ref{eq:ARG2}) with the 
bosonization result for the TLM
given in Eq.\ (\ref{eq:specbos}).
First of all, for $p = 0$ we
obtain from Eq.\ (\ref{eq:ARG2})
 \begin{equation}
 A ( \alpha k_F , \omega ) = \Lambda_0^{- \eta } \frac{\eta}{2} | \omega |^{\eta -1 }
 \; , 
\end{equation}
which corresponds to Eq.\ (\ref{eq:sf:qneq0}) and agrees exactly with the result (\ref{eq:specbos}) obtained via 
bosonization.
Thus, at least for $p=0$, our truncation
of the exact RG flow equations is reliable.
Moreover,  Eq.\ (\ref{eq:ARG2}) exhibits the same scaling
behavior $ A ( \alpha k_F + \alpha s p , s \omega )
 = s^{\eta -1 } A ( \alpha k_F + \alpha p , \omega )$
as the expression (\ref{eq:specbos}) obtained via bosonization.
For finite $p$, however,  the detailed line-shape of the spectral function
predicted by Eq.\ (\ref{eq:ARG2}) is different from Eq.\ (\ref{eq:specbos}).
In particular, for
$ \omega \rightarrow v_c p  + 0$  the bosonization result (\ref{eq:specbos}) predicts 
 \begin{equation}
 A_{\rm TL} ( k_F + p , \omega ) \sim \Lambda_0^{- \eta } \frac{\eta}{2}
  | 2 v_c p |^{\eta / 2 } | \omega - v_c p |^{ -1 + \eta /2 }
 \; ,
 \label{eq:ATLsing}
 \end{equation}
while our RG calculation yields
 \begin{equation}
 A ( k_F + p , \omega ) \sim \Lambda_0^{- \eta } 2 {\eta}
  | \omega - v_c p |^{ -1 + \eta  }
 \; ,
 \label{eq:ARGsing}
 \end{equation}
which diverges with a different exponent than Eq.\ (\ref{eq:ATLsing}).
Similarly, for
$ \omega \rightarrow - v_c p -0 $ the bosonization result (\ref{eq:specbos}) predicts 
 \begin{equation}
 A_{\rm TL} ( k_F + p , \omega ) \sim \Lambda_0^{- \eta } \frac{\eta}{2}
  | 2 v_c p |^{-1 + \eta / 2 } | \omega + v_c p |^{ \eta /2 }
 \; ,
 \label{eq:ATLsing2}
 \end{equation}
while Eq.\ (\ref{eq:ARG2})  yields
 \begin{equation}
 A ( k_F + p , \omega ) \sim \Lambda_0^{- \eta } 2 {\eta}
  | 2 v_c p |^{-1} | \omega + v_c p |^{  \eta  }
 \; .
 \label{eq:ARGsing2}
 \end{equation}
Note, however, that Eq.\ (\ref{eq:ARGsing}) has been derived 
assuming $ | \epsilon^2 - \tilde{v}^2 q^2 | \gg 1$, which means in terms of
the physical variables $ \big| | \omega|  - v_c |p| \big| \gg  1/ \tau = \Lambda$.
For $\Lambda \rightarrow 0$ this condition is always satisfied unless precisely
$  \omega = \pm v_c p$.
This limiting procedure describes the spectral function
precisely at the critical point, 
because
the length $\xi = v_F / \Lambda$ associated with the infrared cutoff
can be sent to infinity {\it{before}} the limit
$ \omega \rightarrow v_c p$ is taken.
Obviously, there exists another regime where both
$ |\epsilon | = | \omega | \tau$ and $ |q | = | p | \xi$ are large, but where
 \begin{equation}
 \big|  | \epsilon|    - \tilde{v}  | q|  \big|   \lesssim 1
 \; .
 \label{eq:limitnew}
 \end{equation}
This corresponds to first taking the
limit
$ \omega \rightarrow \pm v_c p$, and then removing the infrared cutoff
$\Lambda = 1/ \tau \rightarrow 0$.   In this limit
the spectral function behaves completely
different.
Keeping in mind that in practice the
TLM describes a physical system 
only above some finite energy scale $\Lambda_{\ast}$ 
below which
backscattering becomes important or a crossover to higher dimensionality sets in,
this limiting procedure is physically relevant to describe
quasi $1d$ systems. 
If we naively use our result for the
irreducible self-energy given in
Eq.\ (\ref{eq:gammaresscaling}),
we find with the help of Eq.\ (\ref{eq:fasym})
that for $ \epsilon \rightarrow \tilde{v} q$  and for small $ \eta$ 
 \begin{equation}
 \tilde{\Gamma}^{(2 - \mu,Z,v ) }_{\infty} ( q , i \epsilon )  \approx 
\frac{ | q|^{- \eta } }{2} ( i \epsilon  - \tilde{v} q )
 \; .
 \end{equation}
Since by assumption $ | q | \gg 1$, 
this is a negligible correction to the
bare inverse propagator $ i \epsilon - \tilde{v} q$.
It follows that in this regime
our scaling function is simply
  \begin{equation} 
 \tilde{A}_{ \infty } ( q , \epsilon ) \approx \delta ( 
 \epsilon  - \tilde{v} q )
 \; \; , 
 \label{eq:Adelta}
 \end{equation}
so that
the physical spectral function   
exhibits a small but still finite quasiparticle peak
with a weight of the order of
$Z_{l_\ast} = ( \Lambda_{\ast} / \Lambda_0 )^{\eta}$.
Note that this quasiparticle peak is due to the finite 
crossover energy scale $\Lambda_{\ast}$ and does not occur in the TLM
at zero temperature where the infrared cutoff $\Lambda$ 
can be reduced to zero.
In deriving Eq.\ (\ref{eq:Adelta}) 
we have 
introduced the Dirac sea, 
ignored the non-linearity in the energy dispersion as well as 
other irrelevant couplings, which will  broaden the
$\delta$-function.
In principle all these effects can be systematically taken into account
within our formalism. The calculations are quite tedious and are 
beyond the scope of this work. 
It seems, however, that in the regime (\ref{eq:limitnew})
the behavior of the dynamic scaling function
of a generic quasi $1d$ interacting Fermi system with dominant
forward scattering 
is not correctly described by
the scaling function of the Tomonaga-Luttinger model
$\tilde{A}_{\rm TL} ( q , \epsilon )$ given in Eq.\ (\ref{eq:specbosscale}).

\section{Summary and conclusions}
\setcounter{equation}{0}
\label{sec:summary}

In this work we have
shown how to calculate the momentum- and frequency-dependent
single-particle spectral function 
by means of a particular version of the 
functional RG method.\cite{Kopietz01b}
For simplicity, we have applied this method to 
the exactly solvable TLM.
Our truncation scheme of the exact hierarchy of RG flow equations
consists of the following steps:
 \begin{enumerate}
 \item Calculate the flow of the
relevant and marginal couplings
perturbatively for small interactions and adjust  the initial values such that
the RG flow approaches a fixed point.
\item Calculate  the
momentum- and frequency-dependent 
part of the four-point vertex (which involves an infinite number of irrelevant couplings)
in powers of the marginal part of the four-point vertex.

\item Calculate the anomalous dimension in powers of the marginal part of the
four-point vertex.

\item Substitute the results of steps  2. and 3. into the exact
flow equation of the
two-point vertex and then solve this equation exactly.

\end{enumerate}

The result for the spectral function $ A ( k , \omega )$
 is quite encouraging: it has the correct scaling properties, and
 for $k = \pm k_F$
agrees with the exact bosonization result.
For finite $k - k_F$ and $\omega$, we have found evidence that the
spectral line shape in the vicinity of the Luttinger liquid fixed point
exhibits some non-universal features.

Our work also shows how non-Fermi liquid behavior
in a strongly correlated Fermi system can be detected
using modern functional RG methods.
Recently several authors have presented numerical studies 
of the one-loop flow equations for the  marginal part of the 
 {\it{unrescaled}} four-point vertex for 
Hubbard models in $2d$.\cite{Salmhofer98,Halboth00}
The marginal part is obtained by setting  all frequencies 
equal to zero and projecting all wave-vectors onto the Fermi surface.
Ignoring ambiguities\cite{Dupuis98} related to 
different orders of limits, the unrescaled part of the four-point vertex
is then approximated by
 \begin{eqnarray}
  \Gamma^{(4)}_{\Lambda} 
 ( {\bf{k}}_1^{\prime} ,  \omega_1^{\prime}, 
 {\bf{k}}_2^{\prime} , \omega_2^{\prime} ; {\bf{k}}_2 ,  \omega_2, 
 {\bf{k}}_1 ,  \omega_1 )
 &  &
 \nonumber
 \\
& & \hspace{-42mm} \approx
 \Gamma^{(4)}_{\Lambda} 
 ( {\bf{k}}_{F1}^{\prime} ,  0, 
 {\bf{k}}_{F2}^{\prime} , 0 ; {\bf{k}}_{F2} , 0, 
 {\bf{k}}_{F1} , 0 )
 \; . \qquad
 \label{eq:fourpointmarg}
 \end{eqnarray}
For the $2d$ Hubbard model 
with next-nearest neighbor hopping
the numerical analysis\cite{Salmhofer98,Halboth00} of the one-loop flow equations
for the set of marginal couplings contained in the
vertices $ \Gamma^{(4)}_{\Lambda} 
 ( {\bf{k}}_{F1}^{\prime} ,  0, 
 {\bf{k}}_{F2}^{\prime} , 0 ; {\bf{k}}_{F2} , 0, 
 {\bf{k}}_{F1} , 0 )$ 
exhibits a runaway flow to strong coupling
at a finite length scale $l_{\ast} = \ln ( \Lambda_0 / \Lambda_{\ast} )$, which is
typically of the order of
$10^2$. The usual interpretation of this runaway flow
is that it signals some instability of the normal metallic state.\cite{Salmhofer98,Halboth00}
In Ref.\ \onlinecite{Kopietz01b} we have pointed out
that this runaway flow might also be an
artifact of the one-loop approximation, and that
the properly renormalized vertices
could remain finite at the two-loop order.
The crucial point is that the renormalized vertices
involve wave-function renormalization factors (see Eq.\ (\ref{eq:Gammarescale})),
which for a strongly correlated system can cancel a possibly  strong enhancement
of the unrescaled vertices.
Our simple $1d$ model allows us to study such a scenario in detail. 
In this case  $ \Gamma^{(4)}_{\Lambda}$ can be parameterized in terms
of a single marginal coupling $g_l$, 
which is defined in analogy to Eqs.\ (\ref{eq:g2def}) and (\ref{eq:gtdef}),
 \begin{eqnarray}
 {\Gamma}_{\Lambda_0 e^{-l} }^{(4)}
  ( \alpha_1^{\prime}  k_F , 0 , \alpha_2^{\prime} k_F , 0; 
 \alpha_2 k_F , 0 , \alpha_1 k_F , 0 )   & &
 \nonumber
 \\
 & & \hspace{-25mm}  = A_{ \alpha_1^{\prime}  \alpha_2^{\prime} ;
 \alpha_2  \alpha_1 }
 {g}_l
  \; . \qquad
 \label{eq:gtdef2}
 \end{eqnarray}
At the two-loop order, the flow equation of $g_l$  for our model 
is,\cite{Busche02}
 \begin{equation}
 \partial_l g_l = \frac{\nu_0^2}{4}\, g_l^3
 \; .
 \end{equation}
This implies
 \begin{equation}
 g_l = \frac{ g_0}{ \sqrt{ 1 - \frac{\nu_0^2}{2}\, g_0^2 l } }
 \; .
 \label{eq:gflow}
 \end{equation}
Obviously, $g_l$ diverges at a finite scale $l_{\ast} = 2 /(\nu_0 g_0)^2$.
However, this divergence and the associated runaway-flow to strong coupling
are unphysical, because the coupling $g_l$ 
in Eq.\ (\ref{eq:gtdef2}) is {\it{not}} the properly renormalized coupling that
can be identified with the usual $g_2$-interaction of the 
TLM. The latter is defined as a model describing the  {\it{fixed point}} of the RG.
Thus, the $g_2$-coupling that appears in the TLM 
should be identified with the {\it{renormalized coupling
at the RG fixed point}},  which is related to the fixed
point value of our rescaled coupling $\tilde{g}_l$,
 \begin{equation}
 \nu_0 g_2 = \lim_{ l \rightarrow \infty} \tilde{g}_l = \lim_{ l \rightarrow \infty}
 \left[ Z_l^2 \nu_0 g_l \right]
 \; ,
 \label{eq:grenflow}
 \end{equation}
see Eqs.\ (\ref{eq:Gammarescale})  and (\ref{eq:gtdef}).
Similar relations between vertex functions and
interaction parameters at the RG fixed point
are well known from Landau Fermi liquid theory.\cite{Lifshitz80}
In our simple model with a linearized energy dispersion the two-loop flow equation
of the rescaled coupling is simply $\partial_l \tilde{g}_l = 0$, so that
the limit $l \rightarrow \infty$ in Eq.\ (\ref{eq:grenflow}) indeed exists.
Note that the vanishing wave-function renormalization factor $Z_l$
exactly compensates the diverging unrescaled coupling $g_l$ such
that the rescaled coupling $\tilde{g}_l$ remains finite and small.
We believe that the above interpretation of the runaway flow
of the RG-flow equations for the marginal part of the
unrescaled vertices $\Gamma^{(4)}_{\Lambda}$
is not only specific to $1d$, where the
RG fixed point does not correspond to a Fermi liquid.
Assuming that in $2d$ the RG fixed point 
corresponds to  a strongly correlated Fermi liquid 
with $ Z_\infty  \ll 1$,
we expect that the unrescaled vertices  
given in Eq.\ (\ref{eq:fourpointmarg}) will flow to a finite but large
value, which numerically might be indistinguishable from a
runaway flow to infinity. At the same time, the
properly rescaled vertices $\tilde{\Gamma}^{(4)}_l$ defined
in Eq.\  (\ref{eq:Gammarescale}) can remain finite and small.

The problem of calculating 
correlation functions of  many-body systems
at finite wave-vectors or frequencies
using RG methods
has not received much attention.  
For classical systems,  functional RG calculations of 
the momentum-dependence of correlation functions 
can be found in the textbook by Ivanchenko and Lisyansky.\cite{Ivanchenko95}
Here we have presented a functional RG calculation
of a momentum- and frequency-dependent correlation function of a non-trivial 
quantum-mechanical many-body system. For a special $2d$ system a similar calculation has recently been performed
by Ferraz,\cite{Ferraz01} who used the field theory RG.
We believe that the method 
described in this work will also be useful 
to study other problems where no exact solutions are available.
For example, the spectral function of
one-dimensional Fermi systems where backscattering or
Umklapp scattering are relevant 
cannot be calculated exactly by means of bosonization or other methods.
Using our RG method, it should be possible to obtain the spectral function
of Luther-Emery liquids even away from the Luther-Emery point.

\begin{acknowledgments}
We thank  A. Ferraz, V. Meden and K. Sch\"{o}nhammer  for discussions.
This work was partially supported by the DFG via Forschergruppe FOR 412.
\end{acknowledgments}
\appendix
\renewcommand{\theequation}{A.\arabic{equation}}
\renewcommand{\thesubsection}{A.\arabic{subsection}}

%


\begin{thebibliography}{99}
%
\bibitem{Solyom79}
J. S\'{o}lyom, Adv. Phys. {\bf{28}}, 201 (1979).
%
\bibitem{Wilson72}
K. G. Wilson, Phys. Rev. Lett. {\bf{26}}, 548 (1972);
K. G. Wilson and J. G. Kogut, Phys. Reports {\bf{12C}}, 75 (1974).
%
\bibitem{Ma76}
S. K. Ma, {\it{Modern Theory of Critical Phenomena}} (Benjamin/Cummings,
Reading, Massachusetts, 1976).
%
\bibitem{Fisher98} 
M. E. Fisher, Rev. Mod. Phys. {\bf{70}},
653 (1998).
%
\bibitem{footnotesolyom}
Note that in Ref.\ \onlinecite{Solyom79}
the RG $\beta$-functions are derived by means of the 
field theory version of the RG, which relies on the 
renormalizability of the model.
%
\bibitem{Emery79}
V. J. Emery, in {\it{Highly Conducting One-Dimensional Solids}},
edited by J. T. Devreese, R. P. Evrard, and V. E. van Doren
(Plenum, New York, 1979).
%
\bibitem{Voit95}
J. Voit, Rep. Prog. Phys. {\bf{58}}, 977 (1995).
%
\bibitem{Luther74}
A. Luther and I. Peschel, Phys. Rev. B {\bf 9}, 2911 (1974).
%
\bibitem{Meden92}
V. Meden and K. Sch\"{o}nhammer, Phys. Rev. B {\bf{46}}, 15753 (1992).
%
\bibitem{Voit93}
J. Voit, Phys. Rev. B {\bf{47}}, 6740 (1993).
%
\bibitem{Meden99}
V. Meden, Phys. Rev. B {\bf{60}}, 4571 (1999); 
Ph.D. thesis, Universit\"{a}t G\"{o}ttingen,
1996.
%
\bibitem{Haldane81}
F. D. M. Haldane, J. Phys. C {\bf{14}}, 2585 (1981).
%
\bibitem{Halperin69}
B. I. Halperin and P. C. Hohenberg, Phys. Rev. {\bf{177}}, 952 (1969).
%
\bibitem{Sachdev99}
S. Sachdev, {\it{Quantum Phase Transitions}}, (Cambridge University Press,
Cambridge, 1999).
%
\bibitem{Orgad00}
D. Orgad, Philos. Mag.  B {\bf{81}}, 375 (2001);  D. Orgad, S. A. Kivelson,
E. W. Carlson, V. J. Emery, X. J. Zhou, and Z. X. Shen,
Phys. Rev. Lett. {\bf{86}}, 4362 (2001); 
E. W. Carlson, D. Orgad, S. A. Kivelson, and V. J. Emery, Phys. Rev. {\bf{B}},
62, 3422 (2000).
%
\bibitem{Ferraz01}
A. Ferraz, cond-mat/0104576
%
\bibitem{Wegner73}
F. J. Wegner and A. Houghton, Phys. Rev. A {\bf{8}}, 401 (1973).
%
\bibitem{Polchinski84}
J. Polchinski, Nucl. Phys. B {\bf{231}}, 269 (1984).
%
\bibitem{Nicoll76}
J. F. Nicoll, T. S. Chang, and H. E. Stanley, Phys. Lett.
A {\bf{57}}, 7 (1976); 
J. F. Nicoll and T. S. Chang, {\it{ibid.}} {\bf{62}}, 287 (1977);
T. S. Chang, D. D. Vvedensky, and J. F. Nicoll,
Phys. Rep. {\bf{217}}, 279 (1992).
%
\bibitem{Wetterich93}
C. Wetterich, Phys. Lett. B {\bf{301}}, 90 (1993).
%
\bibitem{Morris94}
T. R. Morris, Int. J. Mod. Phys. A {\bf{9}}, 2411 (1994).
%
\bibitem{Zanchi96}
D. Zanchi and H. J. Schulz, Phys. Rev. B {\bf{54}}, 9509 (1996);
Z. Phys. B {\bf{103}}, 339 (1997); Phys. Rev. B {\bf{61}}, 13609 (2000).
%
\bibitem{Salmhofer98}
M. Salmhofer, {\it{Renormalization}} (Springer, Berlin, 1998);
C. Honerkamp, M. Salmhofer, N. Furukawa, and T. M. Rice,
Phys. Rev. B {\bf{63}}, 45114 (2001);
M. Salmhofer and C. Honerkamp, Prog. Theor. Phys. 
 {\bf{105}}, 1, (2001); C. Honerkamp, Ph.D. thesis, 
ETH Z\"{u}rich, 2000.
%
\bibitem{Halboth00}
C. J. Halboth and W. Metzner, Phys. Rev. B {\bf{61}}, 4364 (2000);
Phys. Rev. Lett. {\bf{85}}, 5162 (2001);
C. J. Halboth, Ph.D. thesis, Shaker-Verlag, Aachen, 1999.
%
\bibitem{Kopietz01b}
P. Kopietz and T. Busche, Phys. Rev. B {\bf{64}}, 155101 (2001).
%
\bibitem{Blagoev97}
K. B. Blagoev and K. S. Bedell, Phys. Rev. Lett. {\bf{79}}, 1106
(1997); M. Yamanaka, M. Oshikawa, and I. Affleck,
{\it{ibid.}} 1110 (1997).
%
\bibitem{Schulz98}
H. J. Schulz and B. S. Shastry, Phys. Rev. Lett. {\bf{80}}, 1924 (1998).
%
\bibitem{footnote:g}
Note that the parameter $g_2$ 
of the TLM should not be confused with $g_0$ given in Eq.\ (\ref{eq:g2def}); 
the precise 
relation between these couplings will become evident in 
Sec.\ \ref{sec:summary}, Eq.\ (\ref{eq:grenflow}).
%
\bibitem{footnotevariables}
We are using here a different notation than  in our
previous work:\cite{Kopietz01b}
$\xi$ of Ref.\ \onlinecite{Kopietz01b} is now called $\Lambda$, whereas
$\xi$ is now a length, as is customary in the theory of critical
phenomena.  
%
\bibitem{Kopietz01}
P. Kopietz, Nucl. Phys. B {\bf{595}}, 493 (2001).
%
\bibitem{Luttinger60}
J. M. Luttinger, Phys. Rev. {\bf{119}}, 1153 (1960).
%
\bibitem{Shankar94}
R. Shankar, Rev. Mod. Phys. {\bf{66}}, 129 (1994).
%
%
\bibitem{Dupuis98}
The limit $ {\bf{k}}_i \rightarrow {\bf{k}}_{F_i}$ and $\omega_i \rightarrow 0$
is not unique, so that one has to distinguish between different limiting
procedures. See
N. Dupuis and G. Y. Chitov, Phys. Rev. {\bf{54}}, 3040 (1996).
N. Dupuis, Eur. Phys. J. B {\bf{3}}, 315 (1998).
%
\bibitem{Busche02}
T. Busche and P. Kopietz, unpublished.
%
\bibitem{Lifshitz80}
See, for example, E. M. Lifshitz and L. P. Pitaevskii,
{\it{Statistische Physik, Teil 2}}, (Akademie-Verlag, Berlin, 1980),  page 72.
%
\bibitem{Ivanchenko95}
Y. M. Ivanchenko and A. A. Lisyansky,
{\it{Physics of Critical Fluctuations}}, (Springer, New York, 1995).
%
%
%
%
%
%
%
%

%
%
%
%
%
%
%
%
\end{thebibliography}
\end{document}